\documentclass[twocolumn]{aastex631}
\usepackage[version=4]{mhchem}

\shorttitle{Insights on the Formation Conditions of Uranus and Neptune}
\shortauthors{Mousis et al.}
\graphicspath{{./}{figures/}}

\begin{document}

\title{Insights on the Formation Conditions of Uranus and Neptune from their Deep Elemental Compositions}

\correspondingauthor{Olivier Mousis}
\email{olivier.mousis@lam.fr}

\author[0000-0001-5323-6453]{Olivier Mousis}
\affiliation{Aix-Marseille Universit\'e, CNRS, CNES, Institut Origines, LAM, Marseille, France}
\affiliation{Institut Universitaire de France (IUF), France}
\author[0000-0002-3289-2432]{Antoine Schneeberger}
\affiliation{Aix-Marseille Universit\'e, CNRS, CNES, Institut Origines, LAM, Marseille, France}
\author[0000-0002-0649-1192]{Thibault Cavalié}
\affiliation{Laboratoire d'Astrophysique de Bordeaux, Univ. Bordeaux, CNRS, B18N, allée Geoffroy Saint-Hilaire, 33615, Pessac, France}
\affiliation{LESIA, Observatoire de Paris, Université PSL, CNRS, Sorbonne Université, Université Paris Cité, 5 place Jules Janssen, 92195 Meudon, France}
\author[0000-0001-8397-3315]{Kathleen E. Mandt}
\affiliation{NASA Goddard Space Flight Center, Greenbelt, MD 20771, USA}
\author[0000-0003-2279-4131]{Artyom Aguichine}
\affiliation{Department of Astronomy and Astrophysics, University of California, Santa Cruz, CA, USA}
\author[0000-0003-2279-4131]{Jonathan I. Lunine}
\affiliation{Department of Astronomy, Cornell University, Ithaca, NY, USA}
\author[0000-0002-8719-7867]{Tom Benest Couzinou}
\affiliation{Aix-Marseille Universit\'e, CNRS, CNES, Institut Origines, LAM, Marseille, France}
\author[0000-0001-9275-0156]{Vincent Hue}
\affiliation{Aix-Marseille Universit\'e, CNRS, CNES, Institut Origines, LAM, Marseille, France}
\author[0000-0002-9171-2702]{Rapha\"el Moreno}
\affiliation{LESIA, Observatoire de Paris, Université PSL, CNRS, Sorbonne Université, Université Paris Cité, 5 place Jules Janssen, 92195 Meudon, France}



\begin{abstract}
This study, placed in the context of the preparation for the Uranus Orbiter Probe mission, aims to predict the bulk volatile compositions of Uranus and Neptune. Using a protoplanetary disk model, it examines the evolution of trace species through vapor and solid transport as dust and pebbles. Due to the high carbon abundance found in their envelopes, the two planets are postulated to have formed at the carbon monoxide iceline within the protosolar nebula. The time evolution of the abundances of the major volatile species at the location of the CO iceline is then calculated to derive the abundance ratios of the corresponding key elements, including the heavy noble gases, in the feeding zones of Uranus and Neptune. Supersolar metallicity in their envelopes likely results from accreting solids in these zones. Two types of solids are considered: pure condensates ({\it Case 1}) and a mixture of pure condensates and clathrates ({\it Case 2}). The model, calibrated to observed carbon enrichments, predicts deep compositions. In {\it Case 1}, argon is deeply depleted, while nitrogen, oxygen, krypton, phosphorus, sulfur, and xenon are significantly enriched relative to their protosolar abundances in the two planets. {\it Case 2} predicts significant enrichments for all species, including argon, relative to their protosolar abundances. Consequently, {\it Case 1} predicts near-zero Ar/Kr or Ar/Xe ratios, while {\it Case 2} suggests these ratios are 0.1 and 0.5--1 times their protosolar ratios. Both cases predict a bulk sulfur--to--nitrogen ratio consistent with atmospheric measurements.
\end{abstract}



\section{Introduction} 
\label{sec:sec1}

Uranus and Neptune form a distinct and largely unexplored category of planets within our solar system. In recent years, these two {ice giants} have received increased attention from the scientific community, primarily due to the prevalence of exoplanets with similar masses and sizes, accounting for over half of the more than 5500 exoplanets discovered to date \citep{De20}. Despite their importance, {the vast distances to} Uranus and Neptune have posed formidable challenges to their comprehensive study, with only limited data collected during the Voyager 2 flybys in 1986 and 1989 \citep{li87,Ty86,Sm86,Sm89,Li92,St89}. Consequently, our understanding of these distant worlds relies heavily on remote sensing techniques employed by Earth-based observatories and space telescopes \citep{Ca14,Fe13,Fl10,Fl14,Ir18,Ir19,Ir21,ka09,ka11,Le15,Mo17,Or14a,Or14b,sr14,sr19,Te13,Te21,Te22}. Unfortunately, despite these advances, remote observations are not sufficient to provide direct, unambiguous measurements of the vertical atmospheric structure, composition, and cloud properties of Uranus and Neptune.

The U.S. Decadal Survey has recommended the Uranus Orbiter and Probe (UOP) as the highest priority flagship mission for the next decade\footnote{https://nap.nationalacademies.org/resource/26522/interactive/}. This mission would send an orbiter to study Uranus' internal structure, along with an atmospheric entry probe to directly investigate the planet's atmosphere. By exploring an ice giant planet like Uranus, this mission aims to provide key insights into the formation and evolution of our solar system.

The present study is placed in the context of the preparation {for the UOP mission} and aims at providing predictions on the bulk volatile compositions of both Uranus and Neptune. A protoplanetary disk model is used, and the evolution of trace species is driven by the transport of the various vapors and solids present in the form of dust and pebbles \citep{Sc23}. Due to the high carbon abundance found in their envelopes \citep{sr19,Ir19}, it is postulated that the two planets formed at the CO iceline within the protosolar nebula (PSN). { This formation location} matches their heavy element compositions, as predicted by interior models, with the observed atmospheric D/H ratios \citep{Fe13,Al14}. The time evolution of the abundances of the major volatile species at the location of the CO iceline is then calculated to derive the abundance ratios of the corresponding key elements, including the heavy noble gases, in the feeding zones of Uranus and Neptune. { The supersolar metallicity of the envelopes of the two ice giants is thought to result from the accretion of solids formed in their feeding zones \citep{Lam14,He20}. Two types of solids formed in the PSN are considered: solids composed of pure condensates and solids composed of a mixture of pure condensates and clathrates. The ratio between the two is determined at each time step and radius of the PSN as a function of the availability of volatiles in vapor form and crystalline water, as well as the equilibrium pressures of the different pure condensates and clathrates. Predictions for the deep compositions of Uranus and Neptune can then be made, assuming no significant compositional gradient in the interiors.}

Section \ref{sec:sec2} provides { an overview} of the volatile transport and disk evolution model used in this study. In section \ref{sec:sec3} we outline why the formation of ice giants near the CO iceline is a credible scenario. { In addition, we provide a comprehensive overview of the assumptions underlying our composition model for the envelopes of Uranus and Neptune. In section \ref{sec:sec4}, we present the results of our simulations, outlining the compositions of solids and vapors, along with projections for elemental abundances deep within Uranus and Neptune.} Finally, section \ref{sec:sec5} is devoted to discussion and conclusions.

\begin{figure*}
\includegraphics[width=18cm]{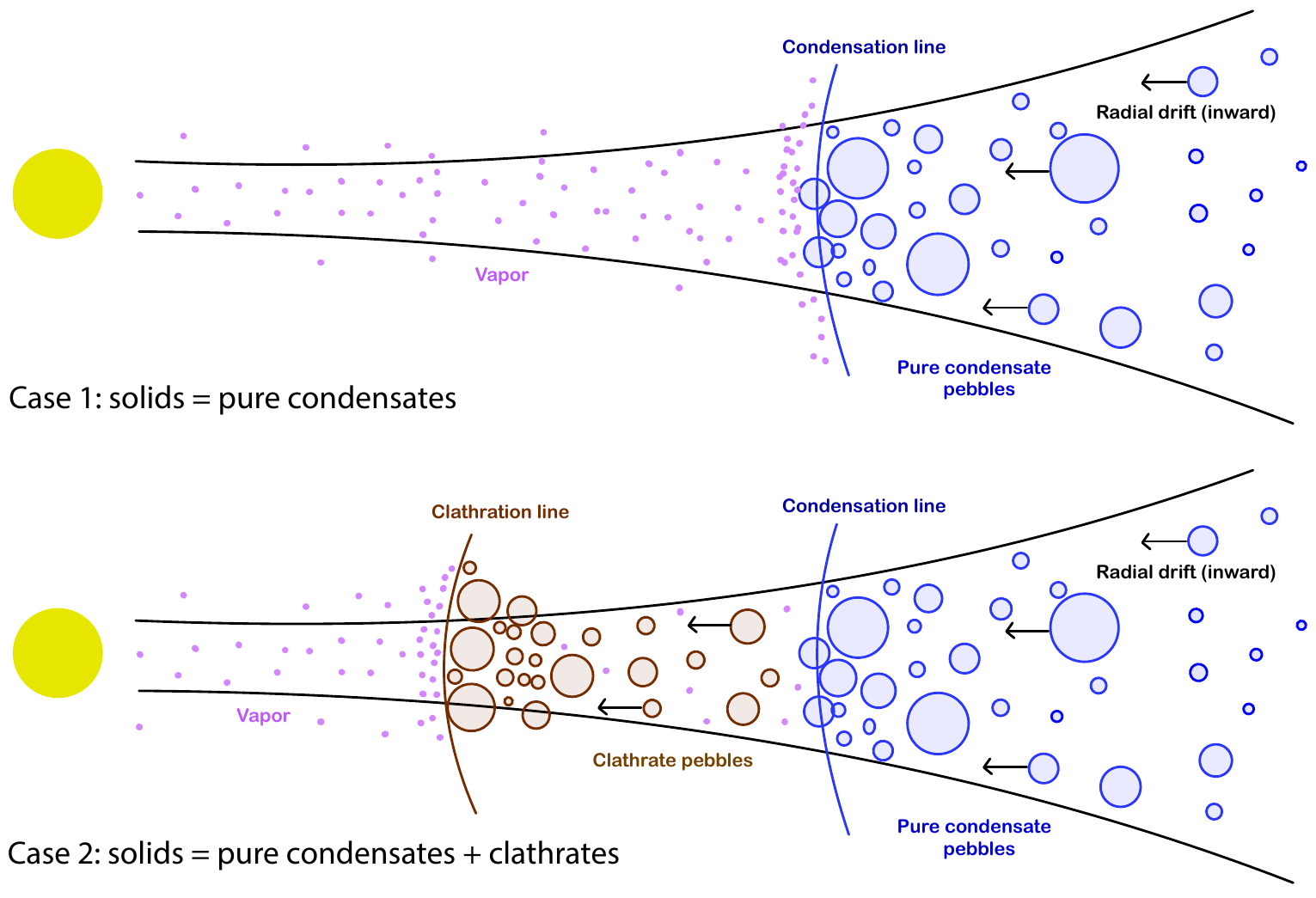}
\caption{Phase changes of volatiles during their transport throughout the PSN. Volatiles are initially delivered as pure condensates in the PSN. { {\it Case 1} and {\it Case 2} correspond to the scenarios where solids can only crystallize as pure condensates, and as pure condensates and clathrates, respectively.} Pure condensate and clathrate pebbles are represented in blue and brown colors for a single species, respectively. Vapor is represented as purple dots. The condensation and clathration lines are represented in blue and brown solid lines, respectively. Once delivered to the disk, the phase of a given species is determined by the relative positions of the corresponding condensation, hydration, or clathration lines. Except for the case of CO$_2$, which vaporises at a higher temperature than its associated clathrate, hydration, or clathration lines of the considered volatiles are closer to the Sun than their respective icelines. Gaseous volatiles condense or become entrapped (depending on the availability of water ice) when diffusing outward of the locations of their condensation, hydration, or clathration lines. Conversely, volatiles condensed or entrapped in grains or pebbles are released in vapor form when drifting inward of their lines. 
} 
\label{fig:fig1}
\end{figure*}

\section{Volatile transport and disk evolution model}
\label{sec:sec2}

{ The disk and transport model utilized in this study is derived from the framework established by \cite{Sc23}. Essentially, this model comprehensively describes the movement of dust particles and vapors within a time-dependent protoplanetary disk, drawing upon methodologies outlined in \cite{ag20} and \cite{mo20}. This study explores two different scenarios: {\it Case 1} corresponds to the scenario where solids can only crystallize as pure condensates, and {\it Case 2} corresponds to the scenario where solids can crystallize as pure condensates and clathrates (see Fig. \ref{fig:fig1} for an overview of the assumptions). The presence of amorphous ice is ruled out under the assumption that volatiles are exclusively delivered as vapors and pure condensates in the PSN. Elements of our model are explicitly described below.}

\subsection{Protoplanetary disk model}

The evolution of the { PSN's surface density} is governed by the classical differential equation \citep{ly74}:

\begin{equation}
\frac{\partial \Sigma_{\mathrm{g}}}{\partial t} = \frac{3}{r} \frac{\partial}{\partial r} \left[ r^{1/2} \frac{\partial}{\partial r} \left( r^{1/2} \Sigma_{\mathrm{g}} \nu \right) \right],
\label{eq:psn}
\end{equation}

\noindent which describes the time evolution of a viscous accretion disk of surface density, $\Sigma_\mathrm{g}$, and dynamical viscosity, $\nu$, assuming hydrostatic equilibrium in the azimuthal direction and invariance in the orbital direction. { This equation can be reformulated into a pair of first-order differential equations that couple the gas surface density $\Sigma_{\mathrm{g}}$ field and the mass accretion rate $\dot{M}$:

\begin{equation}
\begin{cases}
\displaystyle \frac{\partial \Sigma_\mathrm{g}}{\partial t} = \frac{1}{2 \pi r} \frac{\partial \dot{M}}{\partial r} \\
\displaystyle \dot{M} = 3 \pi \Sigma_\mathrm{g} \nu \left( 1 + 2\frac{\partial \ln{\nu \Sigma_\mathrm{g}}}{\partial \ln{r}}\right)
\end{cases} .
\label{eq:resol_sys}
\end{equation}

\noindent  The first equation embodies a mass conservation law, while the subsequent one represents a diffusion equation. The mass accretion rate is delineated as a function of the gas velocity field $v_g$ by the expression $\dot{M}=-2\pi r  \Sigma_{\mathrm{g}} v_{\mathrm{g}}$.}

The viscosity coefficient $\nu$ is computed according to the prescription outlined in \cite{sh73}:

\begin{equation}
\nu = \alpha \frac{c_\mathrm{s}^2}{\Omega_{\mathrm{K}}},
\label{eq:nu}
\end{equation}

\noindent where $\alpha$ is the viscosity coefficient,  $c_\mathrm{s}$ is the sound speed in the PSN, and $\Omega_\mathrm{K}$ is the Keplerian frequency. { The sound speed $c_\mathrm{s}$ is expressed as follows: 

\begin{equation}
c_s = \sqrt{\frac{R T}{\mu_\mathrm{g}}},
\label{eq:cs}
\end{equation}

\noindent  where $\mu_\mathrm{g}$ denotes the mean molecular mass of the gas in the PSN, which is taken here as 2.31 g.mol$^{-1}$, $T$ represents the midplane temperature, and $R$ stands for the ideal gas constant.}

Only viscous heating and the constant irradiation by the local environment of ambient temperature, $T_{\mathrm{amb}}$~=~10 K, are taken into account in our PSN model. Irradiation from the young Sun is neglected because the presence of shadowing is assumed in the outer part of the disk \citep{Oh21}. By doing so, the disk temperature is able to decrease down to the condensation temperature of Ar ($\sim$20 K) in the outer planets region. Such a low disk temperature reached in the outer disk is supported by the {\it in situ} detection of Ar in Comet 67P/Churyumov-Gerasimenko (hereafter 67P/C-G)  by the Rosetta spacecraft \citep{Ba15}. { The temperature profile is determined by summing the production rates from both energy sources, as described by \cite{hu05}:

\begin{equation}
T^4 = \frac{1}{2\sigma_{\mathrm{SB}}} \left( \frac{3}{8}\tau_\mathrm{R} + \frac{1}{2 \tau_\mathrm{P}} \right) \Sigma_\mathrm{g} \nu \Omega_\mathrm{K}^2 + T^4_{\mathrm{amb}},
\label{eq:temp}
\end{equation}

\noindent where $\sigma_{\mathrm{sb}}$ represents the Stefan-Boltzmann constant, and $\tau_\mathrm{R}$ and $\tau_\mathrm{P}$ denote the Rosseland and Planck optical depths, respectively. In this context, we assume $\tau_\mathrm{P} = 2.4 \tau_\mathrm{R}$, which corresponds to the opacity attributed to dust grains \citep{na94}. $\tau_\mathrm{R}$ is determined from the Rosseland mean opacity $\kappa_\mathrm{R}$ using the following relation \citep{hu05}:

\begin{equation}
\tau_R = \frac{\Sigma_g \kappa_R}{2}.
\label{eq:kappaR}
\end{equation}

\noindent  $\kappa_\mathrm{R}$ is calculated using a series of power laws expressed as $\kappa_\mathrm{R} = \kappa_0 \rho^a T^b$, where $\rho$ represents the gas density at the midplane, and $\kappa_0$, $a$, and $b$ are constants determined through fitting to observational data across various opacity regimes \citep{be94}.}

\begin{figure}
\center
\includegraphics[width=\columnwidth]{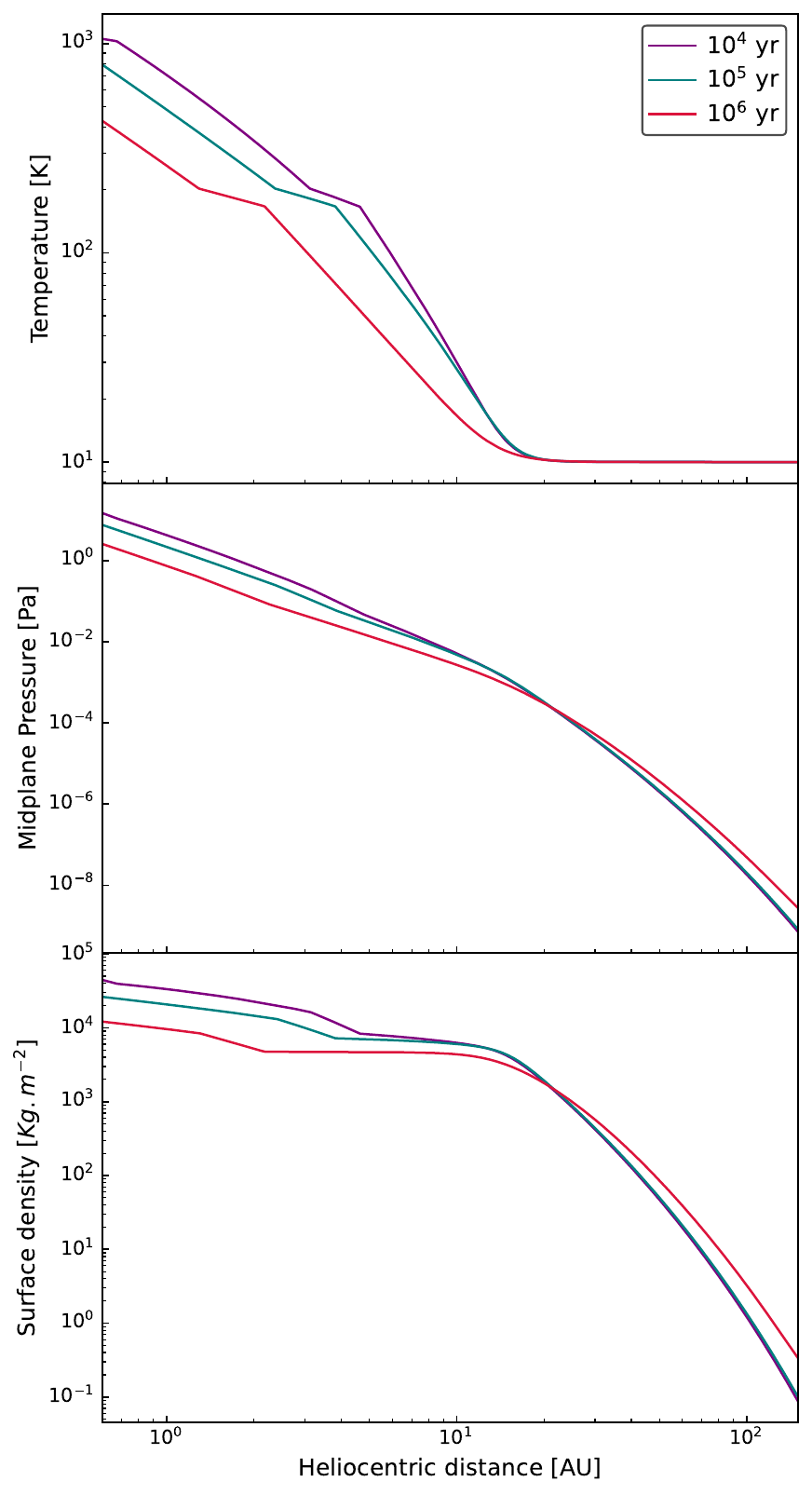}
\caption{From top to bottom: profiles of the disk midplane temperature, pressure and surface density calculated at $t$ = 10$^4$, 10$^5$, and 10$^6$ yr as a function of heliocentric distance, assuming $\alpha$ = $10^{-3}$.} 
\label{fig:fig2}
\end{figure}

{ The initial state of the model is computed based on the self-similar solution derived by \cite{ly74}:

\begin{equation}
\Sigma_\mathrm{g} \nu \propto \exp{\left[ -\left( \frac{r}{r_\mathrm{c}} \right)^{2-p} \right]}.
\label{eq:selfsims}
\end{equation}

\noindent Combining Eqs. \ref{eq:selfsims} and \ref{eq:resol_sys}, and assuming $p=\frac{3}{2}$, corresponding to the case of an early disk \citep{ly74}, yields the initial profiles for the dust surface density and mass accretion rate:

\begin{equation}
\begin{cases}
\displaystyle \Sigma_{\mathrm{g},0} = \frac{\dot{M}_{\mathrm{acc},0}}{3 \pi \nu } \exp \left[ - \left( \frac{r}{r_\mathrm{c}} \right)^{0.5} \right] \\
\displaystyle \dot{M}_0 = \dot{M}_{\mathrm{acc},0} \left(  1 - \left( \frac{r}{r_\mathrm{c}} \right)^{0.5}\right) \exp\left[- \left( \frac{r}{r_\mathrm{c}} \right)^{0.5} \right] 
\end{cases},
\label{eq:init}
\end{equation}

\noindent where $r_\mathrm{c}$ denotes the centrifugal radius, and $\dot{M}_{\mathrm{acc},0}$ represents the initial mass accretion rate onto the central star, set to $10^{-7.6}$ $M_{\odot}$.yr$^{-1}$ \citep{ha98}. The total disk mass is assumed to be $0.1M_{\odot}$, with the majority (99\%) of it confined within $\sim$200 AU. Figure~\ref{fig:fig2} represents the thermodynamic profiles of our nominal PSN model assuming $\alpha$ = $10^{-3}$ as in \cite{Sc23}, and at $t$~=~10$^4$, 10$^5$, and 10$^6$ yr of the disk evolution.}

\subsection{Dust dynamics}
\label{sec2.2}

The dust dynamics in our model follows the two-population algorithm developed by \cite{bir12}, which relies on the key idea that the dynamics of dust pebbles of many different sizes can be well approximated by the dynamics of only { two populations of particles with different sizes (pebble and dust)}. The first group corresponds to a population of small grains with constant size of 0.1 $\mu$m. The second group represents a population of particles of pebble size (a few centimeters), which depends on the characteristics of the flow.

{ Within the disk, pebbles undergo growth through mutual sticking, governed by the following law:

\begin{equation}
a_1(t) = a_0 \exp{\left( \frac{t}{\tau_{\mathrm{growth}}} \right)},
\end{equation}

\noindent where $\tau_{\mathrm{growth}}$ represents the growth timescale, defined as:

\begin{equation}
 \tau_{\mathrm{grow}} = \frac{4 \Sigma_\mathrm{g}}{\sqrt{3}\epsilon_\mathrm{g} \Sigma_\mathrm{b} \Omega_\mathrm{K}},
\end{equation}

\noindent where $\Sigma_{\mathrm{b}}$ denotes the total solid surface density, and $\epsilon_\mathrm{g}$ signifies the dust growth efficiency through mutual sticking, set to 0.5 \citep{la14}. Subsequently, we calculate the Stokes number of the pebbles as a function of their sizes, following the methodology outlined in \cite{jo14}:

\begin{equation}
\mathrm{St} = 
\begin{cases}
\displaystyle \sqrt{2 \pi} \frac{a_1 \rho_\mathrm{b}}{\Sigma_\mathrm{g}} \text{    If      } a_1 \leq \frac{9}{4} \lambda \\ 
\displaystyle \frac{8}{9} \frac{a_1^2 \rho_b c_\mathrm{s}}{\Sigma_\mathrm{g} \nu} \text{    If      } a_1 \geq \frac{9}{4} \lambda
\end{cases}.
\label{eq:stokes}
\end{equation}

\noindent The first and second cases correspond to the Epstein and Stokes regimes, respectively. The boundary between these regimes is determined by the gas mean free path $\lambda = \sqrt{\pi/2}  \cdot \nu / c_\mathrm{s}$, calculated by equating the terms in Eq. (\ref{eq:stokes}). Here, $\rho_{\mathrm{b}}$ represents the mean bulk density of the pebbles:

\begin{equation}
\rho_{\mathrm{b}} = \frac {\sum_i \Sigma_{\mathrm{b},i} \rho_{\mathrm{b},i} } { \sum_i \Sigma_{\mathrm{b},i} },
\end{equation}

\noindent computed as the average of each species' bulk density $\rho_{\mathrm{b},i}$, weighted by their respective solid surface density $\Sigma_{\mathrm{b},i}$. In the following, we assume all pebbles and grains have a density of 1g.cm$^{-2}$ in the PSN, regardless of their compositions.

Observations suggest that disks are abundant in small dust particles, with fragmentation being a dominant process \citep{wi11}. Building upon this observation, our approach incorporates fragmentation and radial drift as the primary growth-limiting mechanisms. These mechanisms impose an upper limit on the highest Stokes number attainable by particles. The first constraint arises from fragmentation, occurring when the relative speed between two pebbles due to turbulent motion exceeds a velocity threshold $u_\mathrm{f}$. This upper limit, described by \cite{bir12}, is:

\begin{equation}
\mathrm{St}_{\mathrm{frag}} = f_{\mathrm{f}} \frac{1}{3 \alpha} \frac{u_\mathrm{f}^2}{c_\mathrm{s}^2},
\end{equation}

\noindent where $u_\mathrm{f}$ and  $f_{\mathrm{f}}$ are set to 10 m.$s^{-1}$ and 0.37, respectively.

Another constraint on dust growth arises from the drift velocities of various pebbles. When pebbles drift faster than they grow, it imposes an additional upper limit on the Stokes number \citep{bir12}:

\begin{equation}
\mathrm{St}_{\mathrm{drift}} = f_{\mathrm{d}} \frac{\Sigma_\mathrm{b} v_\mathrm{K}^2}{\Sigma_\mathrm{g} c_\mathrm{s}^2} \left| \frac{\mathrm{d} \ln P}{\mathrm{d} \ln r} \right|^{-1},
\label{eq:stdrift}
\end{equation}

\noindent where $P$ represents the disk midplane pressure, $v_\mathrm{K}$ denotes the Keplerian velocity, and $f_{\mathrm{d}}$ is set to 0.55.

When dust grains drift at high velocities and collide with other particles in their path, they may undergo fragmentation. This introduces a third upper limit for the Stokes number \citep{bir12}, expressed as: 

\begin{equation}
\mathrm{St}_\mathrm{df} =\frac{1}{1-N} \frac{u_{\mathrm{f}} v_{\mathrm{K}}}{c_{\mathrm{s}}} \left( \frac{\mathrm{d} P}{\mathrm{d} r} \right)^{-1},
\end{equation}

\noindent where the factor $N=0.5$ accounts for the observation that only larger grains tend to fragment upon collision.

In the algorithm, all limiting Stokes numbers are computed and compared with the Stokes number derived from Eq. \ref{eq:stokes}. At each time step, the smallest Stokes number found in this comparison becomes the reference Stokes number which, in turn, sets a new value for the representative size $a_1$ for the large population. The representative size of the small population is always $a_0$, and their Stokes number is always computed in the Epstein regime. 

Among the three size-limiting mechanisms, if particle drift is the most limiting one ($\mathrm{St}_\mathrm{drift} = \min \left(\mathrm{St}_\mathrm{frag},\mathrm{St}_\mathrm{drift},\mathrm{St}_\mathrm{df}\right)$), then the fraction of the mass contained in the large population is $f_\mathrm{m}~=~0.97$. Otherwise, $f_\mathrm{m}$ is set to 0.75 \citep{bir12}. The mean grain size $\bar{a}$ is then given by: 

\begin{equation}
\bar{a} = f_\mathrm{m} a_1 + (1 - f_\mathrm{m})~a_0.
\end{equation}
}


\subsection{Trace species evolution model}
\label{sec:Trace species evolution model}

{ Trace species are accounted for in three distinct forms within our model: vapors, pure condensates, and those entrapped in clathrate. Each of these forms is assigned a distinct surface density denoted by $\Sigma_{\mathrm{v,i}}$, $\Sigma_{\mathrm{p,i}}$, and $\Sigma_{\mathrm{c,i}}$ for species $i$ in the vapor, pure condensate and clathrate/hydrate phases, respectively. Their temporal and radial evolution is governed by the advection-diffusion equation \citep{bir12,de17}:

\begin{equation}
\frac{\partial \Sigma_i}{\partial t} + \frac{1}{r} \frac{\partial}{\partial r} \left[ r \left( \Sigma_i v_i - D_i \Sigma_\mathrm{g} \frac{\partial}{\partial r} \left( \frac{\Sigma_i}{\Sigma_\mathrm{g}} \right) \right) \right] - \dot{Q}_i = 0,
\label{eq:advdiff}
\end{equation}

\noindent where $D_i$ denotes the diffusion coefficient, $v_i$ represents the radial speed, and $\dot{Q}_i$ is a source/sink term that accommodates phase changes, with positive/negative values indicating creation/loss of matter.

Regarding vapor surface densities, we assume $D_i = D_\mathrm{g}$ and $v_i = v_\mathrm{g}$, as vapors are closely coupled to the PSN gas and evolve similarly. Following \cite{sh73}, the gas diffusivity $D_\mathrm{g}$ equals the viscosity $\nu$, and the gas velocity is:

\begin{equation}
v_\mathrm{g} = - \frac{ \dot{M}_{\mathrm{acc}} }{2 \pi r \Sigma_\mathrm{g}}.
\end{equation}

At each time and location, we assume that dust particles form from a mixture of all available solids. Consequently, species trapped in hydrates and amorphous ice possess the same diffusion coefficient $D_s$ and radial velocity $v_s$ as the pure condensates. Following the approach of \cite{bir12}, the diffusion coefficient of solids is determined by:

\begin{equation}
D_\mathrm{s} = \frac{D_\mathrm{g}}{1 + \mathrm{St}^2}.
\end{equation}

\noindent The dust radial velocity is represented as the sum of gas drag and drift velocities, as described by \cite{bir12}:

\begin{equation}
v_\mathrm{s} = \frac{1}{1 + \mathrm{St}^2}v_\mathrm{g} + \frac{2 \mathrm{St}}{1 + \mathrm{St}^2}v_{\mathrm{drift}},
\end{equation}

\noindent where the drift velocity is given by \cite{we97}:

\begin{equation}
v_{\mathrm{drift}} = \frac{c_\mathrm{s}^2}{v_\mathrm{K}} \frac{\mathrm{d}\ln P}{\mathrm{d} \ln r}.
\end{equation}
 
The diffusion coefficient and radial velocity of solids are computed separately for the small and large populations, corresponding to particles of sizes $a_0$ and $a_1$ respectively. Subsequently, the diffusion coefficient $D_{\mathrm{s}}$ and radial velocity $v_{\mathrm{s}}$ employed to evolve surface densities of solids are determined as mass-averaged diffusivities and velocities of the small and large populations, as outlined by \cite{bir12}:}

\begin{equation}
\begin{cases}
\displaystyle v_{\mathrm{s}}= f_\mathrm{m} v_{\mathrm{d},a_0} + (1-f_\mathrm{m}) v_{\mathrm{d},a_1}.\\
\displaystyle D_{\mathrm{s}}= f_\mathrm{m} D_{\mathrm{d},a_0} + (1-f_\mathrm{m}) D_{\mathrm{d},a_1}.
\end{cases}
\end{equation}

\subsection{Sources and sinks of trace species}
\label{sources}

{ We adopt the methodology outlined in \cite{ag20} for illustrating the sources and sinks for both solid and vapor phases of the different species. If the partial pressure of a species $i$'s pure condensate falls below its equilibrium pressure $P_{\mathrm{eq},i}$, it undergoes sublimation. This sublimation process introduces a sink term for pure condensates within each time step $\Delta t$, as detailed in \cite{Dr17}:

\begin{equation}
\Dot{Q}_{\mathrm{p},i} =  - \min \left( \sqrt{\frac{8 \pi \mu_i}{RT}} \frac{3}{\pi \bar{a} \bar{\rho}} P_{\mathrm{eq},i} \Sigma_{\mathrm{p},i} ; \frac{\Sigma_{\mathrm{p},i}}{\Delta t} \right).
\label{eq:ev}
\end{equation}

\noindent Here, $\mu_i$ represents the molar mass of species $i$, $\bar{\rho}$ stands for the mean bulk density of grains, $P_{\mathrm{eq},i}$ denotes the equilibrium pressure, and $\bar{a}$ signifies the mean size of the grains. The latter part of this condition ensures that the quantity of pure condensate sublimated does not exceed the available amount. Details concerning the equilibrium curves of pure condensates can be found in the work of \cite{Sc23}.

Conversely, gas of species $i$ transitions into a pure condensate when its partial pressure surpasses the corresponding equilibrium pressure. This condensation process contributes to source terms for pure condensates, as elaborated in \cite{Dr17}:

\begin{equation}
\Dot{Q}_{\mathrm{p},i} = \min{\left( \left(P_i - P_{\mathrm{eq},i} \right) \frac{2H\mu_i}{RT \Delta t} , \frac{\Sigma_{\mathrm{v},i}}{\Delta t} \right) }, 
\label{eq:cond}
\end{equation}

We have incorporated the possibility of clathrate crystallization in the PSN. Assuming an ample supply of crystalline water, these solids are the first to form during the disk cooling phase due to their higher crystallization temperatures compared to those of corresponding pure condensates. The sole exception to this trend is CO$_2$, which condenses at a higher temperature than its clathrate under nebular conditions. This implies that CO$_2$ pure condensate is considered as the only solid form of CO$_2$ in this model. Furthermore, our model operates under the assumption that NH$_3$ solely exists in its pure condensate form at $t$ = 0 in the PSN. Consequently, NH$_3$ monohydrate cannot form, as its equilibrium curve lies at a lower temperature in the disk compared to its pure condensate. The computation of source/sink terms follows the same methodology as for pure condensates (Eqs. \ref{eq:ev} and \ref{eq:cond}), utilizing equilibrium pressures specific to clathrate hydrates instead of those for pure condensates. Detailed equilibrium curves for various clathrates can be found in the work of \cite{Sc23}. Moreover, the formation of clathrates and hydrates necessitates an adequate presence of crystalline water for trapping. This imposes a constraint on the hydrate source term $\Dot{Q}_{\mathrm{c,i}}$:

\begin{equation}
\Dot{Q}_{\mathrm{c},i}~\Delta t \le \frac{\mu_{i}}{S_i~\mu_{ \ce{H2O}}} \left[ \Sigma_{\mathrm{p},\ce{H2O}} - \sum_k \Sigma_{\mathrm{c},k} \frac{S_k~\mu_{\ce{H2O}}}{\mu_{k}} \right].
\label{eq:cond_cla}
\end{equation}

\noindent Here, $\mu_{i}$ represents the molar mass of species $i$, $\mu_{\ce{H2O}}$ denotes the molar mass of crystalline water, and $S_k$ signifies the stoichiometric ratio between species $k$ and water. This ratio is specifically set to 5.75, and 5.66 for type I and type II clathrates, respectively. 

This expression compares the available amount of crystalline water $\Sigma_{\mathrm{p,water}}$ with the quantity needed to form all clathrates. Throughout the evolution of the disk, this condition is evaluated at each time and location where thermodynamic conditions allow for clathrate formation. It is constructed by assessing the amount of crystalline water available for clathration (as indicated within brackets). If water availability is not a limiting factor, the condition is satisfied. However, if solid water is insufficient, a new hydrate source term $(\Dot Q\mathrm{_c})$ is derived from Eq. \ref{eq:cond_cla}. 

In all scenarios, a prioritization mechanism is employed based on the disparity between $P_i$ and $P_{\mathrm{eq,i}}$. When conditions conducive to clathrate formation are met, the term $P_i - P_{\mathrm{eq,i}}$ yields a positive value. Priority for trapping is accorded to the species with the highest difference value.




From the combined rates of condensation and crystallization, the overall sink and source term for vapor can be derived as follows:}

\begin{equation}
\Dot{Q}_\mathrm{v, \it i} = -\Dot{Q}_{\mathrm{p,\it i}} - \Dot{Q}_{\mathrm{c,\it i}}.
\end{equation}

{ The initial PSN gas phase composition is given in Table~\ref{tab:Abundance}. The mixing ratios are identical to those adopted by \cite{Sc23} in the PSN, but the protosolar elemental abundances have been updated by using the data from \cite{lo21}. We assume that half of C is sequestrated in refractory matter \citep{Be15}. The other half of \ce{C} is distributed between \ce{CO}, \ce{CO2}, or \ce{CH4}, and the remaining \ce{O} forms \ce{H2O}. \ce{CO}:\ce{CO2}:\ce{CH4} and \ce{N2}:\ce{NH3} molecular ratios are assumed to be 10:4:1 and 1:1 in the PSN gas phase, respectively. The \ce{CO}:\ce{CO2} and \ce{CO}:\ce{CH4} ratios are derived from Rosetta measurements of comet 67P/C-G \citep{mou14,ro15}. \ce{N2}:\ce{NH3} is assumed to be~1:1, a value predicted by thermochemical models taking into account catalytic effects of Fe grains on the kinetics of N$_2$ to NH$_3$ conversion in the PSN \citep{Fe00,mou09a}. Sulfur is assumed to be half in \ce{H2S} form and half in refractory sulfide components \citep{pa05}.}

\begin{table}[h]
\centering
\caption{Initial gas phase abundances of the considered species.}
\begin{tabular}{@{}cccc@{}}
\hline
\hline           
\smallskip
Trace species   	&  (X/H$_2$)$_\odot$    		         & Trace species 	          &  (X/H$_2$)$_\odot$ 		    \\
\hline           
\smallskip
\ce{H_2O}       		& $8.86 \times 10^{-4}$        	&       \ce{NH_3}       	& $5.81 \times 10^{-5} $       \\
\ce{CO}         		& $2.42 \times 10^{-4}$        	&       \ce{PH_3}       	& $6.62 \times 10^{-7} $ 		\\
\ce{CO_2}       		& $9.68 \times 10^{-5}$        	&       \ce{Ar}        	    & $7.78 \times 10^{-6} $       \\
\ce{CH_4}       		& $2.42 \times 10^{-5}$        	&       \ce{Kr}         	& $4.08 \times 10^{-9} $  		\\
\ce{H_2S}       		& $1.74 \times 10^{-5} $       	&       \ce{Xe}         	& $4.38 \times 10^{-10} $      \\
\ce{N_2}        		& $5.81 \times 10^{-5} $       	&                           &                               \\
\hline           
\end{tabular}
\label{tab:Abundance}
\end{table}

\section{Ice giants formation at the location of the CO iceline}
\label{sec:sec3}

With values reaching roughly a hundred times the protosolar value (see Sec. \ref{sec:sec4}), Uranus and Neptune exhibit the most pronounced carbon enrichments among the four giants { in our solar system \citep{Mo22}.} This implies that { carbon was a major ingredient in the composition} of the vapors and solids from which the two ice giants formed. The formation of Uranus and Neptune at the CO iceline has been previously proposed by \cite{Al14}. { This scenario is based on the presence of significant amounts of carbon-rich solids at the CO iceline location}, allowing the reconciliation of the observed D/H ratio in Uranus and Neptune's envelopes with the amount of heavy elements predicted by interior models. { In this region of the PSN, CO abundance surpasses that of water, owing to its higher surface density concentrated at its iceline position \citep{Al14}. The accretion process of volatile material by the two ice giants appears to favor carbon--rich ices over water ice. This discrepancy could shed light on why their measured D/H ratios ($\sim$4.1--4.4 $\times$ 10$^{-5}$) lie between the protosolar value observed in Jupiter ($\sim$2 $\times$ 10$^{-5}$) and the higher D-rich values observed in comets ($\sim$1.5--6 $\times$ 10$^{-4}$) \citep{Fe13}. A scenario where a higher fraction of CO is accreted alongside a smaller fraction of cometary ice might align with the observed ratios \citep{Al14}.} Past models struggled to reconcile the observed D/H ratio in Uranus and Neptune with building blocks exhibiting cometary values, { primarily because they assumed primordial water as the primary form of accreted solids \citep{Fe13}}. 

\begin{figure}
\center
\includegraphics[width=\columnwidth]{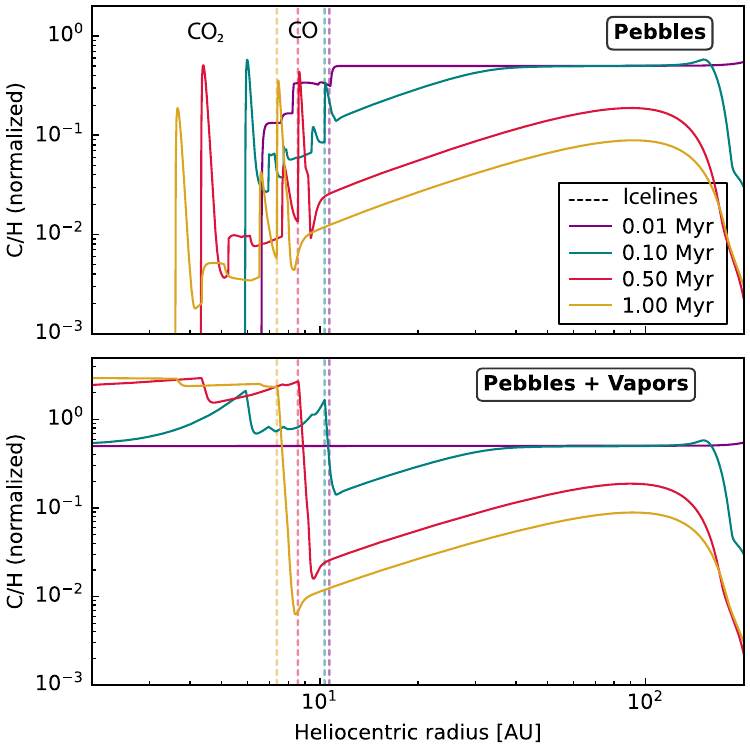}
\caption{Radial profiles of the C/H ratio relative to its protosolar abundance (defined by the enrichment factor $f$), calculated as a function of time in the PSN and in {\it Case 2}, assuming $\alpha$~=~ 10$^{-3}$. Top and bottom panels correspond to the C/H ratio calculated in pebbles and in pebbles + vapors, respectively. Vertical dashed lines indicate CO iceline locations across PSN evolution {(each color corresponds to a different epoch)}.} 
\label{fig:fig3}
\end{figure}

{ Figure \ref{fig:fig3} depicts the radial profiles of the C/H ratio relative to its protosolar abundance \citep{lo21}, and calculated in {\it Case 2}. This analysis assumes that the solids formed consist primarily of pure condensates and clathrates within the PSN.} These profiles are calculated over time in the PSN for both solids and a combination of solids and gas. At each epoch under consideration, { the abundances of solids are zero in regions interior to their corresponding icelines. In this study, we define the iceline of a specific species $i$ as the boundary where the solid-phase surface density $\Sigma_{\mathrm{sol},i}$ equals the vapor-phase surface density $\Sigma_{\mathrm{vap},i}$. This boundary effectively separates the region dominated by ice $i$ from the region dominated by vapor. Consequently, the position of the iceline is governed by both the partial pressure of species $i$ and its corresponding equilibrium vapor pressure. { In Fig. \ref{fig:fig3}, the C/H ratio exhibits two distinct enrichment peaks beyond a few thousand years of PSN evolution, located precisely at the CO$_2$ and CO pure condensate icelines.} These two peaks correspond to the outward diffusion of vapors across the iceline boundaries, where they undergo condensation and accumulate as solid grains at these specific locations. While CO is initially more abundant than CO$_2$ in the gas phase of the PSN, the slightly more pronounced enrichment peak of CO$_2$, positioned at a closer heliocentric distance, is attributed to pebbles drifting at higher velocities along their journey inward the disk. This phenomenon leads to a more substantial accumulation of CO$_2$ in this region of the PSN. In this manner, the mass flux of pebbles crossing the CO$_2$ iceline increases, more than compensating for the lower abundance of CO$_2$. The subtle deviations between the two prominent peaks correspond to minor peaks arising from the condensation/vaporization sequences of CH$_4$ clathrate, CO clathrate, as well as CH$_4$ pure condensate, respectively, as distance from the Sun increases. Plots based on {\it Case 1} would be similar to those in Fig. \ref{fig:fig3}, but without these small variations.}

The C/H enrichment profiles highlight the ease of forming carbon-rich planets from both gases and solids at the CO$_2$ or CO icelines. In the following, we presume that Uranus and Neptune originated from vapors, pebbles, and planetesimals located at the CO iceline. { This hypothesis extends the scope for the formation of the four giants between the H$_2$O and CO icelines. Specifically, it has been postulated that Jupiter's formation likely took place in proximity to the water iceline \citep{Dr17}. Moreover, alternative theories suggest that Jupiter might have originated near the N$_2$ iceline, which could explain its supersolar enrichments in nitrogen and heavy noble gases as detected by the Galileo probe \citep{Ob19,Bo19}. These alternative scenarios advocate for a delayed formation (potentially up to 1 Myr in our model -- see Sec. \ref{sec:sec4}) of Uranus and Neptune at the CO iceline. This delay would have been necessary as the two ice giants would have needed to wait for Jupiter (and Saturn) to migrate inward before start forming themselves.

Our model centers on analyzing the composition of solids and vapors that form and/or evolve within the PSN to provide an estimate of the bulk compositions of the envelopes of Uranus and Neptune. Our primary assumption is that the composition of solids accreted by the two ice giants is fixed during the disk phase and does not evolve during their incorporation by the forming planets. Our sole conjecture concerning planetary formation is that the two ice giants have mostly grown from solids, and gases to a lower extent, situated at the CO iceline because it is the most favorable location to form C--rich material, in agreement with the high carbon enrichment observed in the envelopes (see Sec. \ref{sec:sec4}). Our determination is based on the principle that the heavy elements found in the planetary envelopes were largely supplied by solids during the accretion phase and/or, eventually, by subsequent core erosion  \citep{Lam14,He20}. The composition of these solids mirrors that of the grains that crystallized within the planets' feeding zones in the PSN.}

One should note that, even if our model computes the evolution of the composition of dust and pebbles (with sizes in the $\micron$--cm range) through the different icelines of the PSN, it does not exclude the accretion of larger planetesimals by Uranus and Neptune. Since these solids agglomerated from smaller grains/particles via, for example, streaming instability in the PSN \citep{Ru23}, their composition is fixed by those of their constituting particles and pebbles.

\section{Results} 
\label{sec:sec4}

{ This section is comprised of two parts. The first part explores the composition of solids and vapors evolving around the CO iceline in the PSN. The second part focuses on forecasting the overall abundances within the envelopes of Uranus and Neptune.}

\subsection{Composition of solids and vapors at the CO iceline}
\label{Comp}

\begin{figure*}[h]
\center
\includegraphics[angle=0,width=12cm]{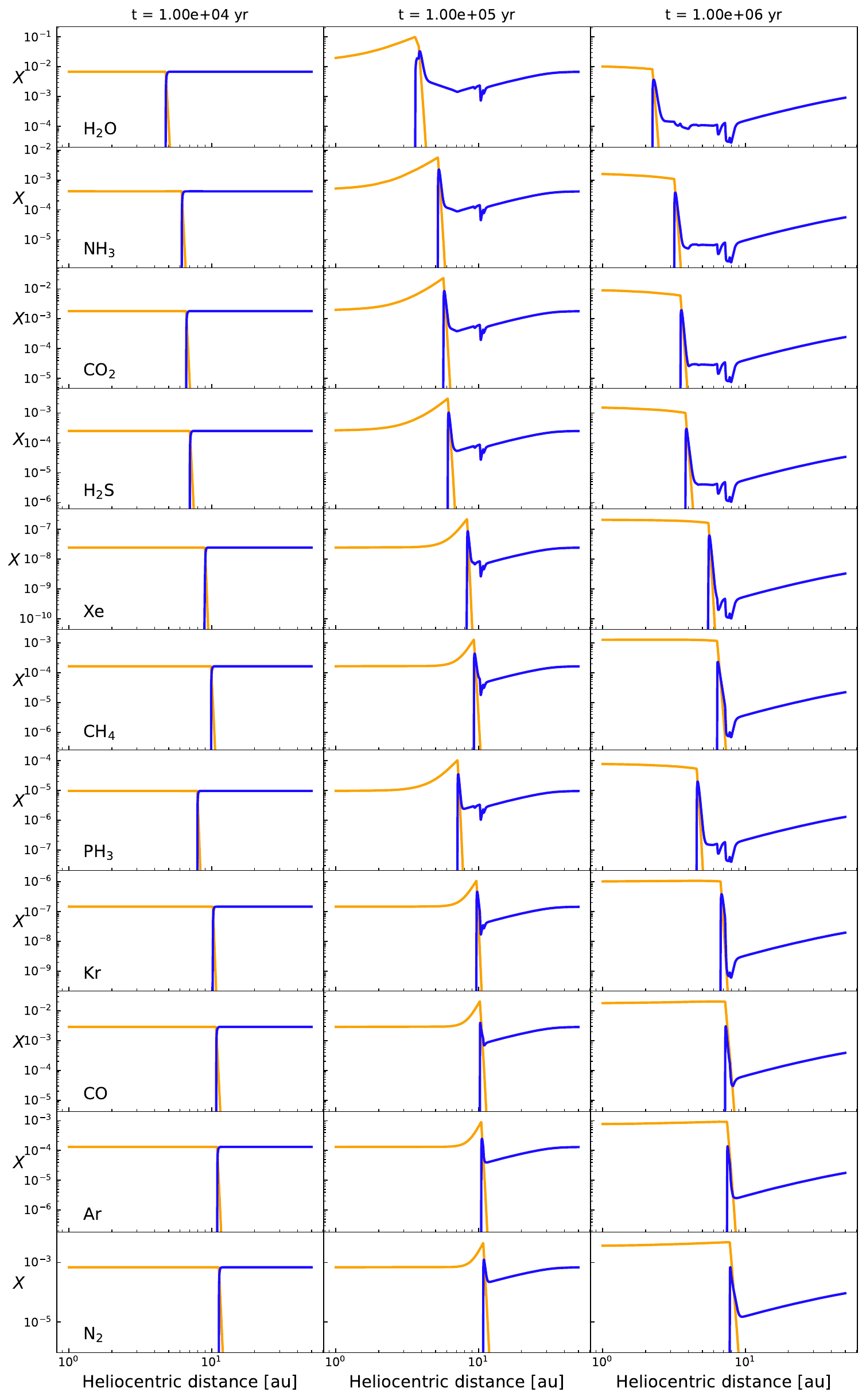}
\caption{ Molar abundance profiles of various volatile species in gaseous (orange lines) and pure condensate (blue lines) forms, calculated at different epochs (10$^4$, 10$^5$, and 10$^6$ yr) of the PSN evolution ({\it Case 1}). The species shown are H$_2$O, NH$_3$, CO$_2$, H$_2$S, Xe, CH$_4$, PH$_3$, Kr, CO, Ar, and N$_2$.}
\label{fig:fig4}
\end{figure*}

\begin{figure*}[h]
\center
\includegraphics[angle=0,width=12cm]{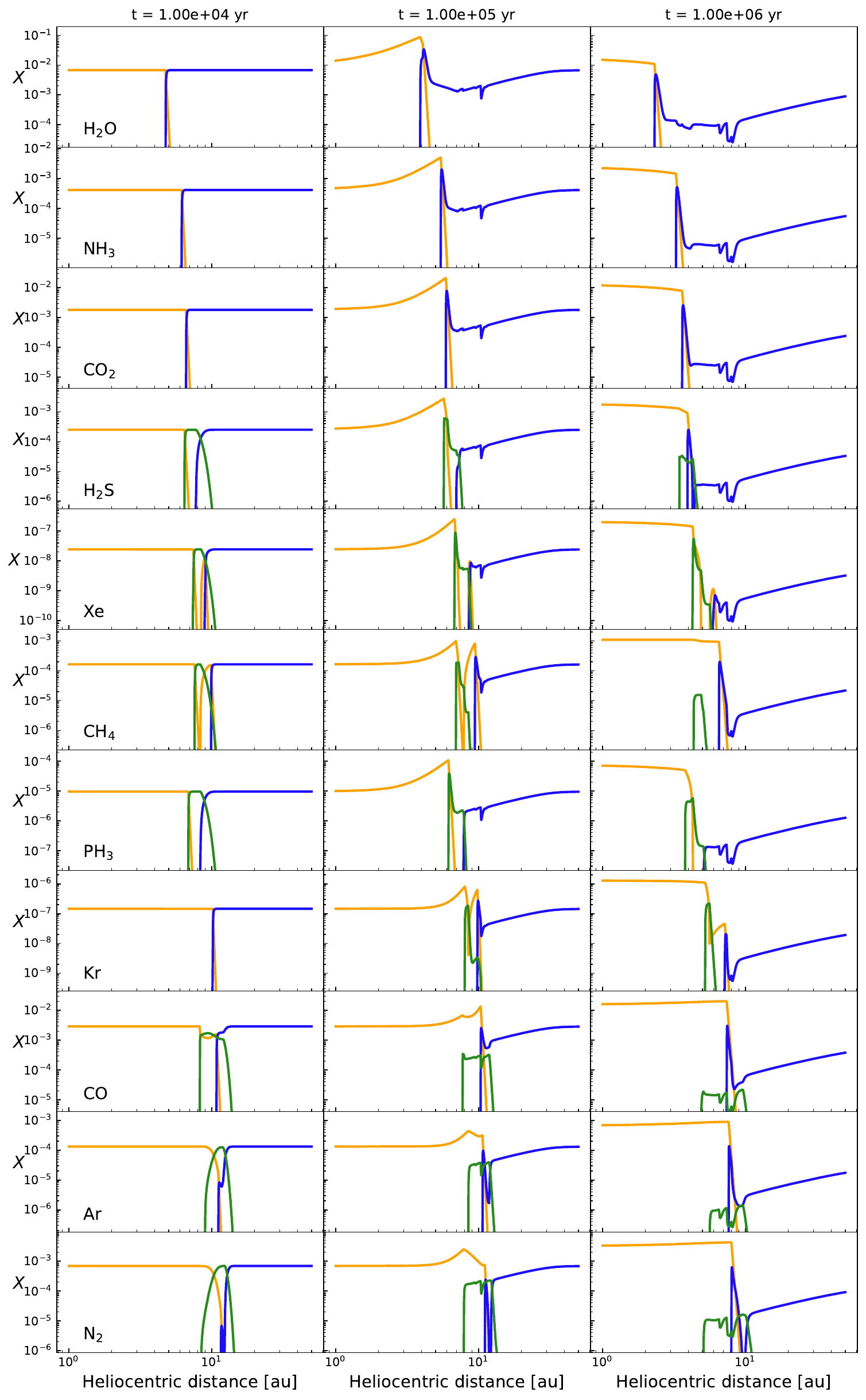}
\caption{ Molar abundance profiles of various volatile species in gaseous (orange lines), clathrate (green lines), and pure condensate (blue lines) forms, calculated at different epochs (10$^4$, 10$^5$, and 10$^6$ yr) of the PSN evolution  ({\it Case 2}). The species shown are H$_2$O, NH$_3$, CO$_2$, H$_2$S, Xe, CH$_4$, PH$_3$, Kr, CO, Ar, and N$_2$.}
\label{fig:fig5}
\end{figure*}

{ Figure \ref{fig:fig4} represents the molar abundance profiles of H$_2$O, NH$_3$, CO$_2$, H$_2$S, Xe, CH$_4$, PH$_3$, Kr, CO, Ar, and N$_2$ in vapors and solids, at different epochs of PSN evolution. The figure illustrates a scenario where vapors exclusively transition into pure condensates during the cooling of the PSN ({\it Case 1}). At $t$ = 10 kyr of disk evolution, the pure condensates start to vaporize as they drift inward in the form of grains, upon penetrating regions interior to approximately 11.2 AU (N$_2$), 10.9 AU (Ar), 10.7 AU (CO), 10.2 AU (Kr), 9.9 AU (PH$_3$), 9.9 AU (CH$_4$), 9.0 AU (Xe), 7.0 AU (H$_2$S), 6.6 AU (CO$_2$), 6.2 AU (NH$_3$), and 3.3 AU (H$_2$O) from the Sun within the PSN. All grains are vaporized once they have drifted inward past the iceline of H$_2$O. At $t$ = 100 kyr, the cooling of the PSN brings the sublimation region of the grains closer to the Sun, typically spanning from approximately $\sim$2.6 to 11.0 AU from the Sun. At $t$~=~1~Myr, the icy grains sublimate at closer heliocentric distances, namely in the 1.4--8.0 AU region from the Sun. 

Figure \ref{fig:fig5} is analogous to Fig. \ref{fig:fig4}, but it includes the possibility that vapors transition into solids either constituted of clathrates or pure condensates ({\it Case 2}). Here, the positions of the icelines of pure condensates align with those of Fig. \ref{fig:fig4}. Comparing the two figures shows that clathrates can persist several AU closer to the Sun compared to their corresponding pure condensates. With the exception of H$_2$O, CO$_2$, and NH$_3$, which exist solely in their pure condensate forms, all other solid grains are completely vaporized upon crossing the clathrate lines. These lines are positioned closer to the Sun with respect to those of the pure condensate forms, approximately at $\sim$8.4 AU (N$_2$),  9.0 AU (Ar), 8.2 AU (CO), 8.5 AU (Kr), 6.9 AU (PH$_3$), 7.6 AU (CH$_4$), 7.4 AU (Xe), and 6.4 AU (H$_2$S) from the Sun at 10 kyr of PSN evolution. The abundance of clathrates in the solid phase is contingent upon the availability of crystalline water, which itself is influenced by the fraction of carbon present in oxidized C-bearing volatiles. With the assumed mixing ratios in the initial PSN gas phase (see Sec. \ref{sources}), clathrates can be as abundant as the corresponding pure condensates at closer heliocentric distances, dominating the composition of solids formed in the vicinity of their icelines.}

\begin{figure*}
\center
\includegraphics[angle=0,width=15cm]{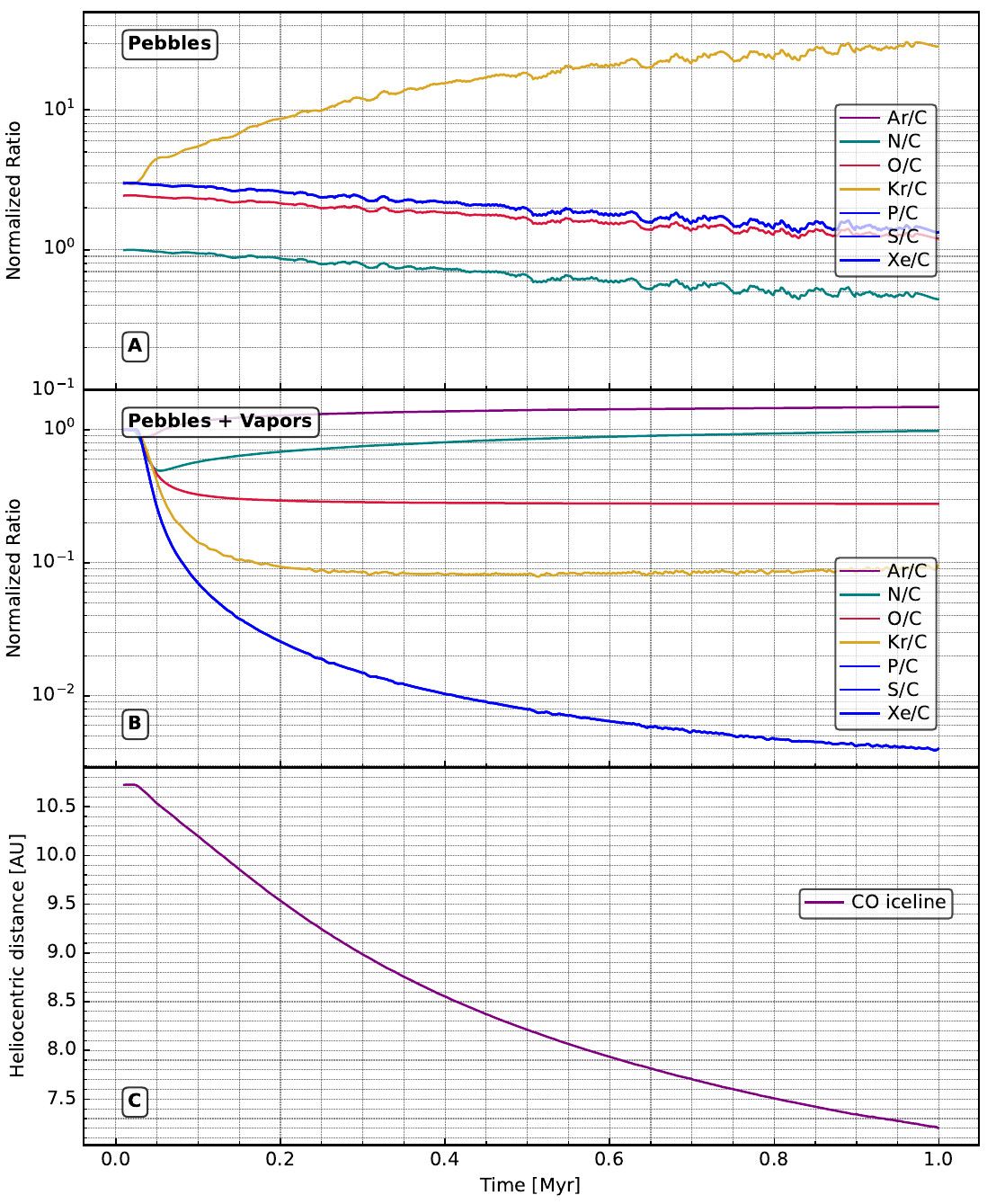}
\caption{ Panel ({\it a}): temporal evolution of elemental ratios (Ar/C, N/C, O/C, Kr/C, P/C, S/C, and Xe/C) calculated in {\it Case 1} for solid condensates at the CO iceline within the PSN. The ratios are adjusted to their protosolar values. The P/C, S/C, and Xe/C ratios appear superimposed in the panel (see text for details). The Ar/C ratio is not represented in the solids because it falls below 10$^{-25}$ (see text). Panel ({\it b}): same elemental ratios as in panel (a), but summed over both solid and vapor phases within the PSN. Panel ({\it c}): temporal evolution of the position of the CO iceline within the PSN.}
\label{fig:fig6}
\end{figure*}

\begin{figure*}
\center
\includegraphics[angle=0,width=15cm]{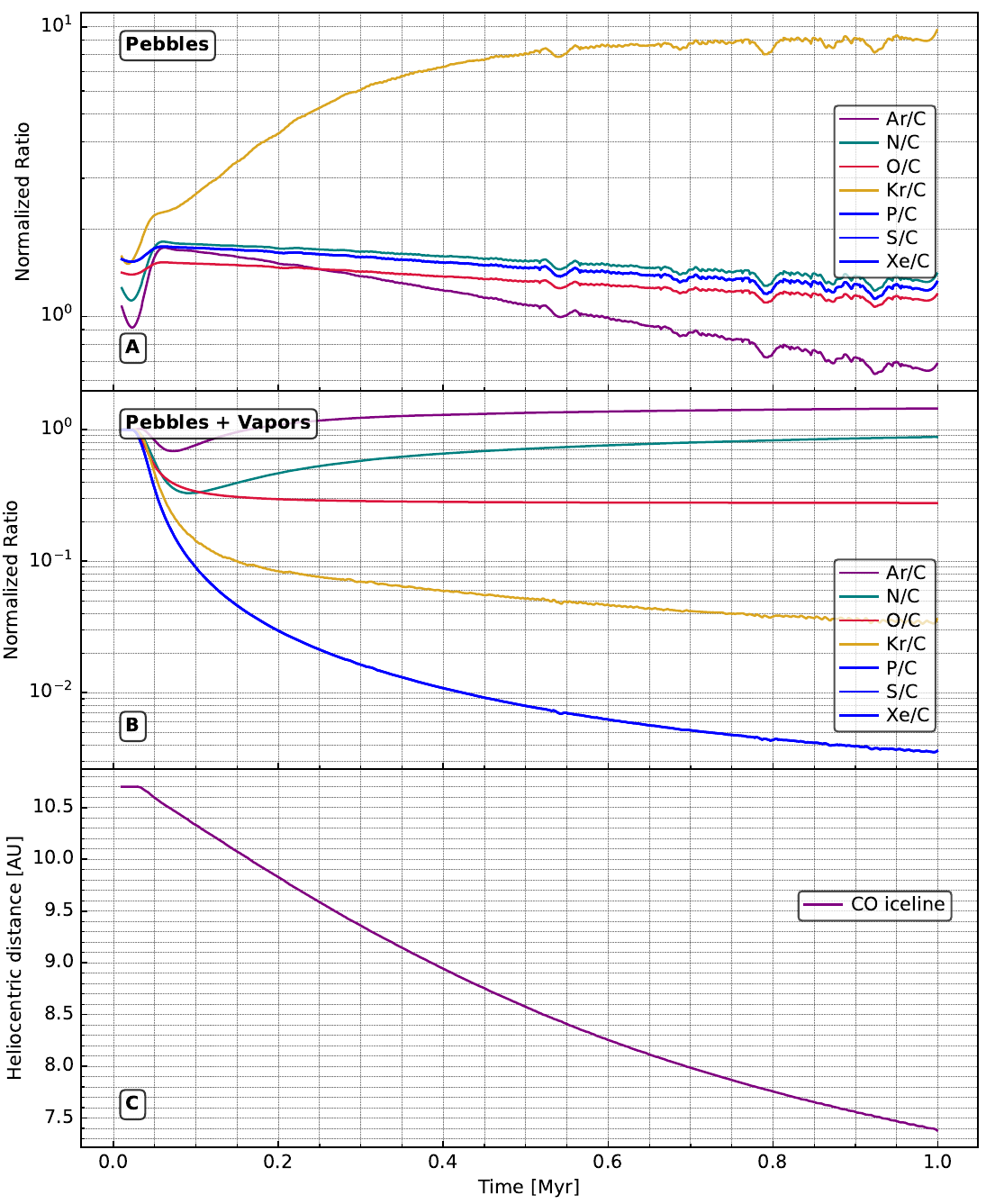}
\caption{ Same as in Fig. \ref{fig:fig6}, but in {\it Case 2}.}
\label{fig:fig7}
\end{figure*}

{ Figures \ref{fig:fig6} and \ref{fig:fig7} depict the temporal evolution of the total elemental ratios (Ar/C, N/C, O/C, Kr/C, P/C, S/C, and Xe/C) relative to their protosolar values in {\it Case 1} and {\it Case 2}, respectively. These ratios are calculated by summing the contributions from both solid and solid+vapor phases at the precise location of the CO iceline in the PSN. The CO iceline gradually migrates inward from approximately 10.7 AU to about 7.2 AU within the first Myr of the PSN evolution.} P/C, S/C, and Xe/C elemental ratios appear superimposed { in the two figures} because all P, S, and Xe elements are locked into single species (H$_2$S, PH$_3$, and Xe) that form pure condensates evolving in a similar manner at the location of the CO iceline. At the beginning, all ratios are protosolar in the disk, which takes several dozen thousands of years of evolution to enable the transport of enough matter and form the different enrichment peaks. 

{ In both {\it Case 1} and {\it Case 2}, when considering the dual contribution of vapors+solids,} the global Ar/C ratio rapidly evolves towards a plateau slightly higher { ($\sim$1.5 times)} than its protosolar ratio because of the significant contribution of Ar vapor at the CO iceline location. { Remarkably, even though only 0.7 times the protosolar ratio after 1 Myr of PSN evolution, the Ar/C ratio is orders of magnitude higher in solids in {\it Case 2} compared to the {\it Case 1} (10$^{-25}$ at the same epoch). This results from the presence of Ar clathrate at the CO iceline, while pure Ar still fails to condense at this location.}

{ In both cases, the global N/C, and O/C ratios decrease down to plateaus reaching $\sim$0.9--1 and 0.3 times their protosolar ratios, respectively. This decrease is primarily due to an increase in the amount of solid CO at its iceline. The N/C ratio in solids is three times higher in {\it Case 2} compared to {\it Case 1}. This difference can be attributed to a significant portion of N$_2$ being sequestered in N$_2$ clathrate in {\it Case 2}, whereas solid N$_2$ is nearly absent at the CO iceline. Similarly, the decrease of the global Kr/C ratio is more pronounced in {\it Case 2} ($\sim$0.04 $\times$ (Kr/C)$_{\odot}$) than in {\it Case 1} ($\sim$0.1 $\times$ (Kr/C)$_{\odot}$) after 1 Myr of PSN evolution. This reduction is attributed to the presence of Kr clathrates closer to the Sun in {\it Case 2}. These Kr clathrates effectively sequester a significant amount of vapor, thereby reducing the portion that can diffuse outward from the disk.

In both {\it Case 1} and {\it Case 2}, the increase in the CO surface density at the location of its iceline is the primary cause of the steep decreases in the global P/C, S/C, and Xe/C ratios to less than 4~$\times$~10$^{-3}$ their protosolar ratios after 1 Myr of PSN evolution. This is because the icelines of PH$_3$, H$_2$S and Xe are located at closer heliocentric distances than that of CO in the PSN.  Consequently, these relative ratios become more sensitive to the sharp increase in the CO abundance.}

\subsection{Predicted bulk elemental abundances}

\begin{figure*}
\includegraphics[angle=0,width=15cm]{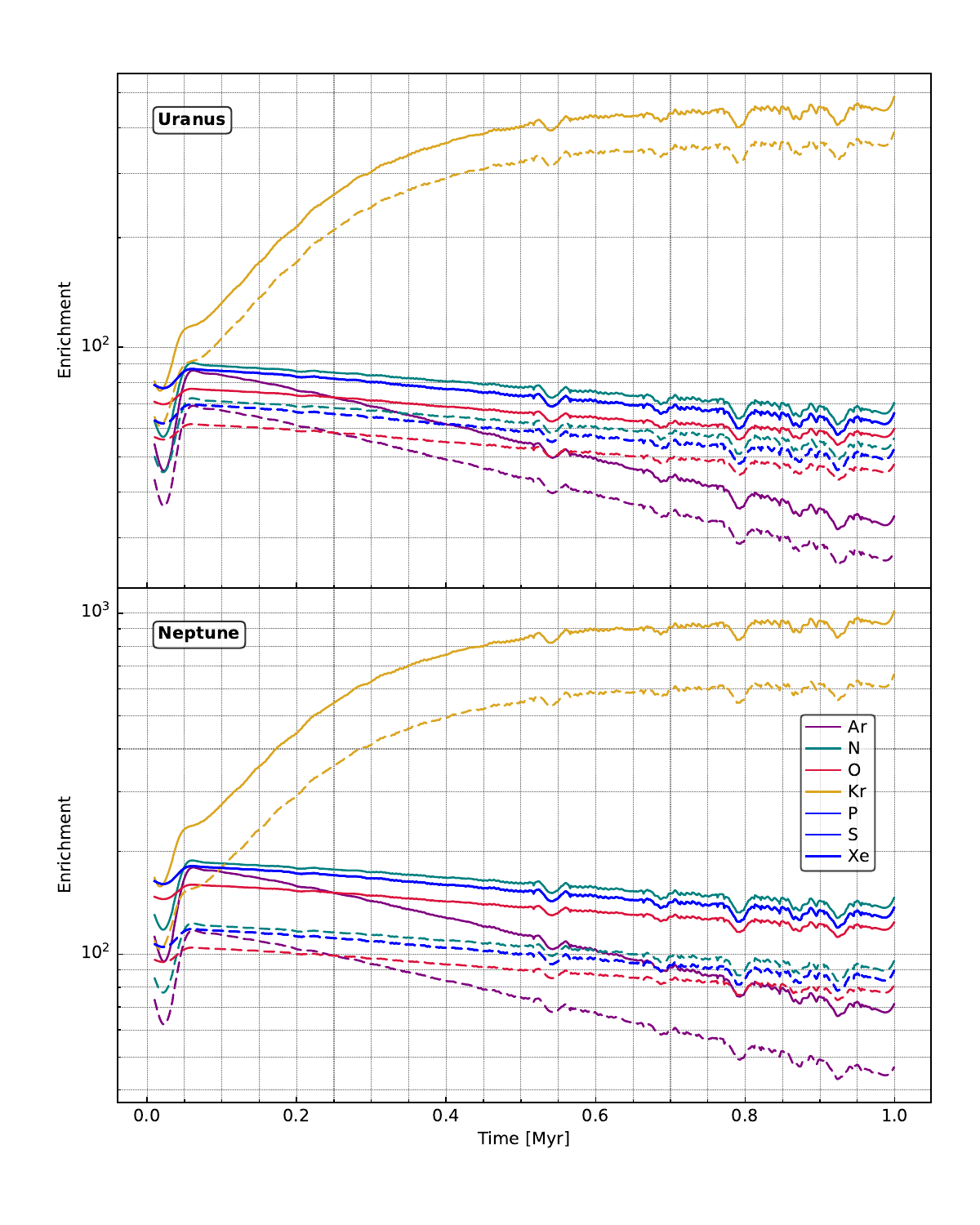}
\caption{ Top panel: elemental enrichments with respect to protosolar abundances predicted in the envelope of Uranus in {\it Case 1}, assuming a C--enrichment ranging between 40 and 50 times the protosolar value. Dashed and solid lines correspond to predictions of the lower and upper values for the elemental enrichments, respectively.  Bottom panel: same as in top panel but for Neptune, assuming a C--enrichment ranging between 68 and 104 times the protosolar value.} 
\label{fig:fig8}
\end{figure*}

\begin{figure*}
\includegraphics[angle=0,width=15cm]{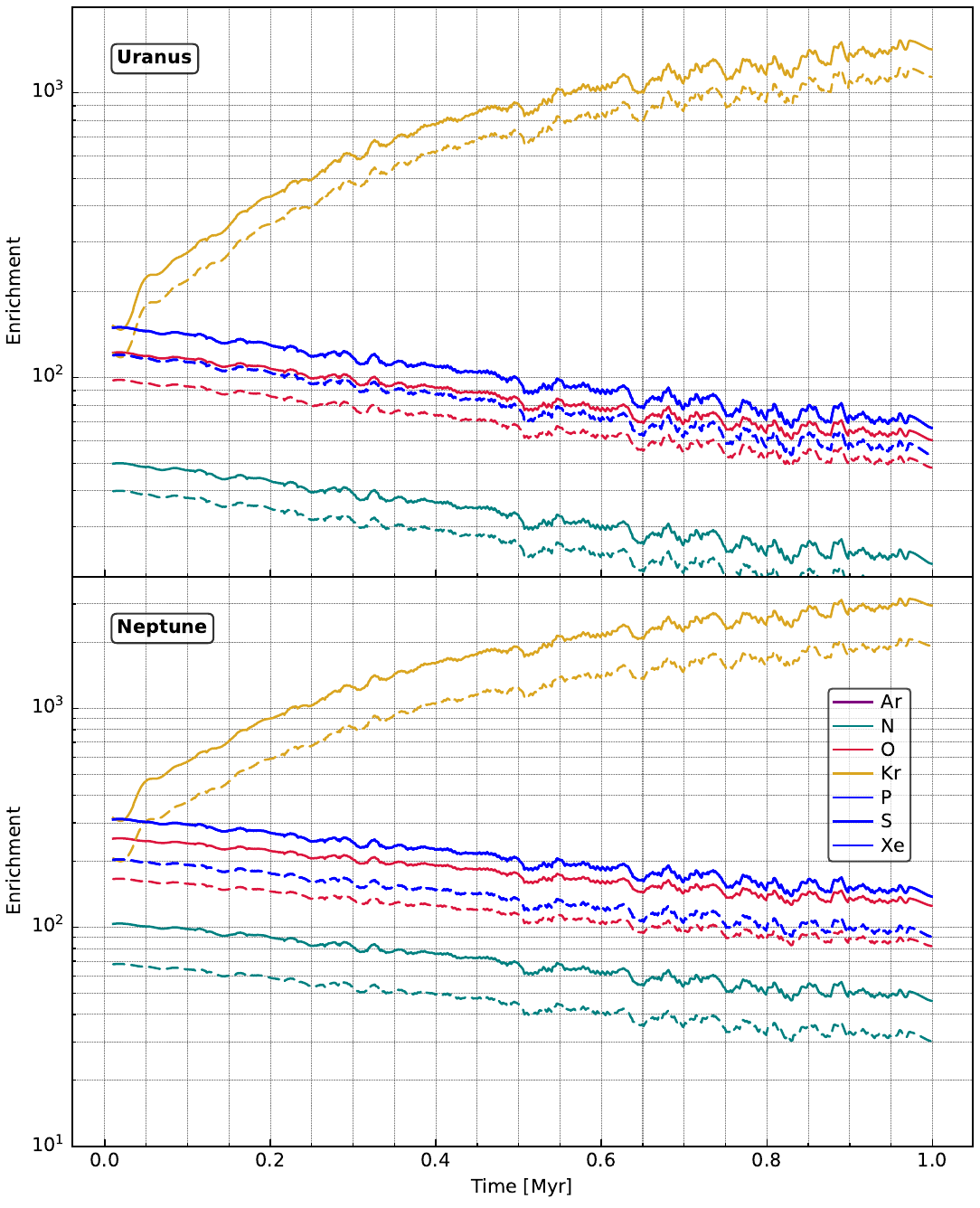}
\caption{ Same as in Fig. \ref{fig:fig8}, but in {\it Case 2}.} 
\label{fig:fig9}
\end{figure*}

{ Figures \ref{fig:fig8} and \ref{fig:fig9} further illustrate the translation of the Ar/C, N/C, O/C, Kr/C, P/C, S/C, and Xe/C ratios into bulk elemental enrichments relative to the protosolar elemental abundances in the envelopes of Uranus and Neptune in {\it Case 1} and {\it Case 2}, respectively. These elemental ratios are also computed as a function of the time evolution of the PSN. To calibrate the model, we use the observed carbon enrichments that have been measured remotely in Uranus and Neptune. Specifically, we multiply the Ar/C, N/C, O/C, Kr/C, P/C, S/C, and Xe/C ratios computed in solids by the measured carbon abundances in the atmospheres of Uranus and Neptune. This approach allows us to infer the overall enrichment levels of other key elements in the envelopes of these ice giant planets, based on their measured carbon enrichment. By tying the other elemental abundances to the measured carbon abundances, we can obtain an estimate of the bulk composition of Uranus and Neptune's atmospheres. In each figure, the predicted bulk abundances in Uranus and Neptune correspond to the scenario where the metallicities of the two planets' envelopes have been primarily supplied by the delivery of solid materials, either from the dilution of their cores or during the accretion process of the forming planets. Table \ref{tab:pred} extracts the predictions for volatile enrichments in the envelopes of Uranus and Neptune from Figures \ref{fig:fig8} and \ref{fig:fig9}, assuming their heavy elements were acquired from the PSN at 0.5 and 1 Myr of its evolution. Most of the volatile abundance predictions reach steady-state conditions after 1 Myr of PSN evolution.}

In the case of Uranus, a CH$_4$ abundance is set between 2.4 and 3$\%$ \citep{sr19}, corresponding to an enrichment factor that is $\sim$40--50 times the protosolar value \citep{lo21}, and adopting the He abundance derived from Voyager 2 observations \citep{co87}. In the case of Neptune, the CH$_4$ abundance is ranged between 4 and 6$\%$ \citep{Ir19}, leading to an enrichment factor that is $\sim$68--104 times the protosolar value \citep{lo21}, and adopting the He abundance derived from ISO long-wavelength observations \citep{Bu03}.

{ Assuming that the heavy elements present in Uranus's envelope were supplied solely by solids formed in the PSN at 0.5 Myr (or 1 Myr) of evolution, the model predicts the following  enrichment or impoverishment in volatiles in {\it Case 1}:

\begin{itemize}
\item Ar is absent from the envelopes or severely depleted,
\item N, O, and Kr are enriched by factors of 26.2--32.8 (17.7--22.2), 67.2--84.0 (48.2--60.3), and 729.7--912.2 (1135.9--1419.8), respectively, with respect to their protosolar values,
\item P, S, and Xe are all enriched by a similar factor of 78.7--98.4 (53.2--66.6), with respect to their protosolar values.
\end{itemize}

\noindent The depletion of Ar arises from the fact that it condenses at a greater heliocentric distance compared to the CO iceline. Although the predicted volatile abundances from solids formed at 1 Myr of PSN evolution are slightly different from the 0.5 Myr case, they remain qualitatively similar. The key distinctions are the lower enrichment factors for N, O, P, S, and Xe at 1 Myr, while Kr exhibits a higher enrichment factor compared to the 0.5 Myr case.

Under the same assumptions regarding formation conditions, our remarks on volatile abundances are also valid for Neptune. Apart from the strong depletion of Ar, which is consistent with both cases, the higher enrichments of other volatiles correspond to the calibration of their abundances to the higher measured carbon enrichment in Neptune's atmosphere.

Assuming that the heavy elements present in Uranus's envelope were supplied solely by solids formed in the PSN at 0.5 Myr (or 1 Myr) of evolution, the model predicts the following  enrichment or impoverishment in volatiles in {\it Case 2}:

\begin{itemize}
\item Ar, N, O, and Kr are enriched by factors of 44.0--55.0 (27.4--34.3), 62.3--77.8 (56.2--70.3), 53.0--66.2 (47.7--59.6), and 323.8--404.7 (389.2--486.6), respectively, with respect to their protosolar values,
\item P, S, and Xe are all enriched by a similar factor of 59.0--73.7 (52.6--65.8), with respect to their protosolar values.
\end{itemize}

\noindent Clathrates of various volatile species form at locations closer than the respective ice lines of their pure condensates. This alters the predicted volatile enrichments in Uranus compared to the simple {\it Case 1} scenario of pure condensates only. A key difference is the presence of Ar clathrates at the CO iceline. This engenders a supersolar Ar/C ratio at that location, contrasting with the significant Ar depletion noted in {\it Case 1} due to the lack of an Ar solid reservoir in the PSN. Similar remarks regarding the role of clathrates in trapping volatiles can be formulated for the volatile enrichments in Neptune.

\begin{table*}
\centering
\caption{Volatile enrichments predicted in Uranus and Neptune at 0.5 and 1 Myr of PSN evolution.} 
\begin{tabular}{lllcccccccc}
\hline
\hline           
\smallskip
\smallskip
Case	 		& Epoch		& Value 		& Ar			& N		& O		& Kr		& P		& S		& Xe		\\
\hline           
\smallskip
			&			&			& \multicolumn{6}{c}{Uranus}										\\
{\it Case 1}	& 0.5 Myr		& Min		& 0.0 		& 26.2	& 67.2	& 729.7	& 78.7	& 78.7	& 78.7	\\
			& 			& Max 		& 0.0			& 32.8	& 84.0	& 912.2	& 98.4	& 98.4	& 98.4	\\
			& 1 Myr		& Min 		& 0.0			& 17.7	& 48.2	& 1135.9	& 53.2	& 53.2	& 53.2	\\
			&			& Max 		& 0.0			& 22.2	& 60.3	& 1419.8	& 66.6	& 66.6	& 66.6	\\
			&			&			& \multicolumn{6}{c}{Neptune}\\
 			& 0.5 Myr		& Min		& 0.0			& 44.6	& 114.3	& 1240.6	& 133.8	& 133.8	& 133.8	\\
			& 			& Max 		& 0.0			& 68.2	& 174.8	& 1897.3	& 204.6	& 204.6	& 204.6	\\
			& 1 Myr		& Min 		& 0.0			& 30.2	& 82.0	& 1931.0	& 90.5	& 90.5	& 90.5	\\
		 	&			& Max 		& 0.0			& 46.1	& 125.4	& 2953.3	& 138.4	& 138.4	& 138.4	\\
\hline           
\smallskip
			&			&			& \multicolumn{6}{c}{Uranus}\\
{\it Case 2} 	& 0.5 Myr		& Min		& 44.0		& 62.3	& 53.0	& 323.8	& 59.0	& 59.0	& 59.0	\\
			& 			& Max 		& 55.0		& 77.8	& 66.2	& 404.7	& 73.7	& 73.7	& 73.7	\\
			& 1 Myr		& Min 		& 27.4		& 56.2	& 47.7	& 389.2	& 52.6	& 52.6	& 52.6	\\
			&			& Max 		& 34.3		& 70.3	& 59.6	& 486.6	& 65.8	& 65.8	& 65.8	\\
			&			&			& \multicolumn{6}{c}{Neptune}\\
 			& 0.5 Myr		& Min		& 74.7		& 105.8	& 90.0	& 550.4	& 100.3	& 100.3	& 100.3	\\
			& 			& Max 		& 114.3		& 161.9	& 137.7	& 841.8	& 153.3	& 153.3	& 153.3	\\
			& 1 Myr		& Min 		& 46.7		& 95.6	& 81.0	& 661.7	& 89.4	& 89.4	& 89.4	\\
			&			& Max 		& 71.4		& 146.2	& 123.0	& 1012.0	& 136.8	& 136.8	& 136.8	\\
\hline           
\end{tabular}
\label{tab:pred}
\end{table*}
    
\section{Discussion and conclusion} 
\label{sec:sec5}

{ In this work, we have investigated the key compositional tracers that would be associated with the formation of Uranus and Neptune along the CO iceline in the PSN, assuming the composition of their building blocks was shaped by the radial transport of particles, dust, and vapors in the disk. A protoplanetary disk and transport model has been used to study the evolution of volatile species throughout the PSN, and predict the bulk compositions of Uranus and Neptune. The model tracks the condensation/sublimation of pure condensates and formation of clathrates for different volatile species. Our determination is based on the principle that the heavy elements found in the planetary envelopes were essentially supplied by these solids during the accretion phase and/or by subsequent core erosion. The composition of the solids mirrors that of the grains that crystallized within the planets' feeding zones in the PSN. Two cases have been considered in our simulations: {\it Case 1} assumes that the solids evolving in the PSN were only formed from pure condensates, and {\it Case 2} assumes that these solids could consist of condensates, clathrates, or both. The abundances of the various volatile species in gas, pure condensate, and clathrate forms are calculated as a function of time and heliocentric distance with our disk model.}

Both {\it Case 1} and {\it Case 2} predict significant enrichments for most species in the envelopes of Uranus and Neptune. However, the key difference between the two cases is the inclusion of clathrates in {\it Case 2}, which significantly affects the predicted Ar abundance. Whereas {\it Case 1} predicts a significant depletion of this noble gas in the envelopes due to its absence in the solids accreted by the two planets, in {\it Case 2} significant amounts of Ar should be trapped in the clathrates that participate in the composition of the solids considered. As a result, supersolar enrichments of Ar are predicted in the envelopes of Uranus and Neptune (see Table \ref{tab:pred}). Interestingly, the predictions of our model are not very sensitive to the formation epoch of the two ice giants. Although the predicted volatile abundances of solids formed at 1 Myr of PSN evolution are slightly different from the 0.5 Myr case, they remain close.

Any variation in the formation sequence of Uranus and Neptune does not affect the calculations of the composition of the accreted bodies, provided that both planets formed at the CO iceline. However, if one of these planets accreted its heavy elements beyond or below the CO iceline, it would be more challenging to explain the observed carbon enrichment in its atmosphere. This is because the ratio of carbon--to--hydrogen is higher in solids formed in this region of the PSN, as shown in Fig. \ref{fig:fig3}. However, our approach is consistent with planets migrating inward once they have accreted the majority of their heavy elements. 

While future in situ measurements promise to refine our understanding of the locations and timing of the formation of the two planets, current uncertainty in carbon measurements used in our model is significant. The error bars associated with these measurements are too wide to effectively explore potentially different chemical pathways between the two planets.}

The various elements which are carried by condensible species can only be measured below their respective cloud decks. Following the cloud vertical structure presented in \cite{At20}, C, S and N, and finally O would then require a probe to reach a pressure level of $\sim$2 bar, 30--50 bar, and several hundreds bar, respectively. Because of these important depths, it is desirable that a probe be able to measure secondary carrier species which do not condensate to probe these elements. Such measurements of the abundances of, \textit{e.g.}, CO, N$_2$ for O and N, require the support of thermochemical models to constrain the deep abundance of these elements \citep{Ca20}. In the context of probe measurements, the prediction of the { Ar/Kr or Ar/Xe} abundance ratios is useful for the assessment of our model because, so far, there is no known differentiation mechanism between these noble gases as long as they remain in the vapor phase. { {\it Case 1} predicts that the Ar/Kr or Ar/Xe ratios in Uranus and Neptune are close to zero, while in {\it Case 2} they should be on the order of 0.1 and 0.5--1 times their protosolar ratios in the two planets, respectively. In particular, Xe stands out as the only heavy noble gas capable of condensing in the atmosphere of Uranus, although its cloud level is expected to be in the range of about 0.1 to 0.4 bar \citep{Za23}. However, an entry probe would penetrate below this altitude and reach the layer where the noble gases are thoroughly mixed. Measuring the noble gases in situ would be key to testing our various scenarios, since their abundance ratios would in principle not be affected by any possible compositional heterogeneity, as measured in Jupiter by the Juno mission \citep{He20,Ba21}.}

Assuming our calculations are representative of homogeneous interiors, the derived O/H ratio is in agreement with the values obtained in Uranus from the thermochemical model of \cite{Ca17} and \cite{Ve20}. At Neptune, the O/H is nominally lower than the model results derived from tropospheric CO observations \citep{Mo17}. However, to what extent this value should  be considered as an upper limit for the deep neptunian O/H remains debated, as the tropospheric abundance of CO could be much lower than previously thought \citep{Te19}. New broadband and high spectral resolution observations of CO in Neptune are crucially required. { The current upper limit for PH$_3$ stands at 1.1 ppb, as reported by \cite{Te19}, established within pressure ranges of 0.1 to 1.2 bar, a zone where this compound experiences significant photochemical loss.} It is thus not possible to use this observation as a constraint for comparison with the P/H value resulting from our simulations. The fact that our model predicts more PH$_3$ than the upper limit set by \cite{Te19} confirms the photochemical losses in the upper troposphere. { Beyond 0.5 Myr of simulation, we obtain an S/N ratio of $\sim$15 in {\it Case 1} and $\sim$5 in {\it Case 2}. This is in agreement with} the measurements of \cite{Ir18} and \cite{Ir19}, who find that S/N exceeds unity in Uranus and Neptune. Their detection of H$_2$S at pressures of a few bars suggests a prevalence of H$_2$S over NH$_3$, as indicated by the survival of some H$_2$S after the formation of NH$_4$SH clouds at depths of 30 to 50 bars. 

{ It is important to note that the positions of icelines in our model are contingent upon the PSN temperature/pressure profiles, which themselves rely on a number of factors, including solar irradiation, the chosen viscosity parameter, disk opacity, and others. Consequently, it is crucial to acknowledge the inherent variability in determining the precise locations of icelines. Models that posit the formation of ice giants at greater heliocentric distances, such as within the 12–17 AU range \citep{Ts05}, do not inherently contradict our findings in this regard.

It is also notable that the formation lines for clathrates are situated closer to the Sun compared to the CO iceline. Consequently, the impact of this solid reservoir is reduced at the location of the ice giants, which are assumed to form at the CO iceline within the PSN. If we consider the possibility that Jupiter and Saturn formed from pebbles at heliocentric distances closer to the Sun, the significance of clathrates in their formation processes could be enhanced. This possibility warrants further investigation in the future.}







\begin{thebibliography}{}

\bibitem[Aguichine et al.(2020)]{ag20} Aguichine, A., Mousis, O., Devouard, B., et al.\ 2020, \apj, 901, 97. doi:10.3847/1538-4357/abaf47

\bibitem[Ali-Dib et al.(2014)]{Al14} Ali-Dib, M., Mousis, O., Petit, J.-M., et al.\ 2014, \apj, 793, 9. doi:10.1088/0004-637X/793/1/9


\bibitem[Aguichine et al.(2022)]{ag22} Aguichine, A., Mousis, O., \& Lunine, J.~I.\ 2022, Planet. sci. j. , 3, 141. doi:10.3847/PSJ/ac6bf1


\bibitem[Atreya et al.(2020)]{At20} Atreya, S.~K., Hofstadter, M.~H., In, J.~H., et al.\ 2020, \ssr, 216, 18. doi:10.1007/s11214-020-0640-8

\bibitem[Bailey \& Stevenson(2021)]{Ba21} Bailey, E. \& Stevenson, D.~J.\ 2021, \psj, 2, 64. doi:10.3847/PSJ/abd1e0

\bibitem[Balsiger et al.(2015)]{Ba15} Balsiger, H., Altwegg, K., Bar-Nun, A., et al. \ 2015, Science Advances, 1:e1500377 (2015).

\bibitem[Bell \& Lin(1994)]{be94} Bell, K.~R. \& Lin, D.~N.~C.\ 1994, \apj, 427, 987. doi:10.1086/174206

\bibitem[Bergin et al.(2015)]{Be15} Bergin, E.~A., Blake, G.~A., Ciesla, F., et al.\ 2015, Proceedings of the National Academy of Science, 112, 8965. doi:10.1073/pnas.1500954112

\bibitem[Birnstiel et al.(2012)]{bir12} Birnstiel, T., Klahr, H., \& Ercolano, B.\ 2012, \aap, 539, A148. doi:10.1051/0004-6361/201118136


\bibitem[Bosman et al.(2019)]{Bo19} Bosman, A.~D., Cridland, A.~J., \& Miguel, Y.\ 2019, \aap, 632, L11. doi:10.1051/0004-6361/201936827

\bibitem[Burgdorf et al.(2003)]{Bu03} Burgdorf, M., Orton, G.~S., Davis, G.~R., et al.\ 2003, \icarus, 164, 244. doi:10.1016/S0019-1035(03)00138-6

\bibitem[Cavali{\'e} et al.(2020)]{Ca20} Cavali{\'e}, T., Venot, O., Miguel, Y., et al.\ 2020, \ssr, 216, 58. doi:10.1007/s11214-020-00677-8

\bibitem[Cavali{\'e} et al.(2017)]{Ca17} Cavali{\'e}, T., Venot, O., Selsis, F., et al.\ 2017, \icarus, 291, 1. doi:10.1016/j.icarus.2017.03.015

\bibitem[Cavali{\'e} et al.(2014)]{Ca14} Cavali{\'e}, T., Moreno, R., Lellouch, E., et al.\ 2014, \aap, 562, A33. doi:10.1051/0004-6361/201322297

\bibitem[Conrath et al.(1987)]{co87} Conrath, B., Gautier, D., Hanel, R., et al.\ 1987, \jgr, 92, 15003. doi:10.1029/JA092iA13p15003

\bibitem[Deleuil et al.(2020)]{De20} Deleuil, M., Pollacco, D., Baruteau, C., et al.\ 2020, \ssr, 216, 105. doi:10.1007/s11214-020-00726-2

\bibitem[Desch et al.(2017)]{de17} Desch, S.~J., Estrada, P.~R., Kalyaan, A., et al.\ 2017, \apj, 840, 86. doi:10.3847/1538-4357/aa6bfb


\bibitem[Dra{\.z}kowska \& Alibert(2017)]{Dr17} Dra{\.z}kowska, J. \& Alibert, Y.\ 2017, \aap, 608, A92. doi:10.1051/0004-6361/201731491



\bibitem[Fegley(2000)]{Fe00} Fegley, B.~J.\ 2000, Space Science Reviews, 92, 177

\bibitem[Feuchtgruber et al.(2013)]{Fe13} Feuchtgruber, H., Lellouch, E., Orton, G., et al.\ 2013, \aap, 551, A126. doi:10.1051/0004-6361/201220857

\bibitem[Fletcher et al.(2014)]{Fl14} Fletcher, L.~N., de Pater, I., Orton, G.~S., et al.\ 2014, \icarus, 231, 146. doi:10.1016/j.icarus.2013.11.035

\bibitem[Fletcher et al.(2010)]{Fl10} Fletcher, L.~N., Drossart, P., Burgdorf, M., et al.\ 2010, \aap, 514, A17. doi:10.1051/0004-6361/200913358

\bibitem[Hartmann et al.(1998)]{ha98} Hartmann, L., Calvet, N., Gullbring, E., et al.\ 1998, \apj, 495, 385. doi:10.1086/305277

\bibitem[Helled et al.(2022)]{He22} Helled, R., Stevenson, D.~J., Lunine, J.~I., et al.\ 2022, \icarus, 378, 114937. doi:10.1016/j.icarus.2022.114937
 
\bibitem[Helled \& Fortney(2020)]{He20} Helled, R. \& Fortney, J.~J.\ 2020, Philosophical Transactions of the Royal Society of London Series A, 378, 20190474. doi:10.1098/rsta.2019.0474

\bibitem[Hueso \& Guillot(2005)]{hu05} Hueso, R. \& Guillot, T.\ 2005, \aap, 442, 703. doi:10.1051/0004-6361:20041905

\bibitem[Irwin et al.(2021)]{Ir21} Irwin, P.~G.~J., Dobinson, J., James, A., et al.\ 2021, \icarus, 357, 114277. doi:10.1016/j.icarus.2020.114277

\bibitem[Irwin et al.(2019)]{Ir19} Irwin, P.~G.~J., Toledo, D., Braude, A.~S., et al.\ 2019, \icarus, 331, 69. doi:10.1016/j.icarus.2019.05.011

\bibitem[Irwin et al.(2018)]{Ir18} Irwin, P.~G.~J., Toledo, D., Garland, R., et al.\ 2018, Nature Astronomy, 2, 420. doi:10.1038/s41550-018-0432-1

\bibitem[Johansen et al.(2014)]{jo14} Johansen, A., Blum, J., Tanaka, H., et al.\ 2014, Protostars and Planets VI, 547. doi:10.2458/azu\_uapress\_9780816531240-ch024

\bibitem[Karkoschka \& Tomasko(2009)]{ka09} Karkoschka, E. \& Tomasko, M.\ 2009, \icarus, 202, 287. doi:10.1016/j.icarus.2009.02.010

\bibitem[Karkoschka \& Tomasko(2011)]{ka11} Karkoschka, E. \& Tomasko, M.~G.\ 2011, \icarus, 211, 780. doi:10.1016/j.icarus.2010.08.013

\bibitem[Lambrechts \& Johansen(2014)]{la14} Lambrechts, M. \& Johansen, A.\ 2014, \aap, 572, A107. doi:10.1051/0004-6361/201424343

\bibitem[Lambrechts et al.(2014)]{Lam14} Lambrechts, M., Johansen, A., \& Morbidelli, A.\ 2014, \aap, 572, A35. doi:10.1051/0004-6361/201423814

\bibitem[Le Roy et al.(2015)]{ro15} Le Roy, L., Altwegg, K., Balsiger, H., et al.\ 2015, \aap, 583, A1. doi:10.1051/0004-6361/201526450

\bibitem[Lellouch et al.(2015)]{Le15} Lellouch, E., Moreno, R., Orton, G.~S., et al.\ 2015, \aap, 579, A121. doi:10.1051/0004-6361/201526518

\bibitem[Lindal(1992)]{Li92} Lindal, G.~F.\ 1992, \aj, 103, 967. doi:10.1086/116119


\bibitem[Lindal et al.(1987)]{li87} Lindal, G.~F., Lyons, J.~R., Sweetnam, D.~N., et al.\ 1987, \jgr, 92, 14987. doi:10.1029/JA092iA13p14987

\bibitem[Lodders(2021)]{lo21} Lodders, K.\ 2021, \ssr, 217, 44. doi:10.1007/s11214-021-00825-8

\bibitem[Lynden-Bell \& Pringle(1974)]{ly74} Lynden-Bell, D. \& Pringle, J.~E.\ 1974, \mnras, 168, 603. doi:10.1093/mnras/168.3.603

\bibitem[Moreno et al.(2017)]{Mo17} Moreno, R., Lellouch, E., Cavali{\'e}, T., et al.\ 2017, \aap, 608, L5. doi:10.1051/0004-6361/201731472

\bibitem[Mousis et al.(2022)]{Mo22} Mousis, O., Atkinson, D.~H., Ambrosi, R., et al.\ 2022, Experimental Astronomy, 54, 975. doi:10.1007/s10686-021-09775-z

\bibitem[Mousis et al.(2014)]{mou14} Mousis, O., Lunine, J.~I., Fletcher, L.~N., et al.\ 2014, \apjl, 796, L28. doi:10.1088/2041-8205/796/2/L28

\bibitem[Mousis et al.(2009)]{mou09a} Mousis, O., Lunine, J.~I., Thomas, C., et al.\ 2009, \apj, 691, 1780. doi:10.1088/0004-637X/691/2/1780


\bibitem[Mousis et al.(2020)]{mo20} Mousis, O., Aguichine, A., Helled, R., et al.\ 2020, Philosophical Transactions of the Royal Society of London Series A, 378, 20200107. doi:10.1098/rsta.2020.0107

\bibitem[Nakamoto \& Nakagawa(1994)]{na94} Nakamoto, T. \& Nakagawa, Y.\ 1994, \apj, 421, 640. doi:10.1086/173678

\bibitem[{\"O}berg \& Wordsworth(2019)]{Ob19} {\"O}berg, K.~I. \& Wordsworth, R.\ 2019, \aj, 158, 194. doi:10.3847/1538-3881/ab46a8

\bibitem[Ohno \& Ueda(2021)]{Oh21} Ohno, K. \& Ueda, T.\ 2021, \aap, 651, L2. doi:10.1051/0004-6361/202141169

\bibitem[Orton et al.(2014b)]{Or14b} Orton, G.~S., Moses, J.~I., Fletcher, L.~N., et al.\ 2014b, \icarus, 243, 471. doi:10.1016/j.icarus.2014.07.012

\bibitem[Orton et al.(2014a)]{Or14a} Orton, G.~S., Fletcher, L.~N., Moses, J.~I., et al.\ 2014a, \icarus, 243, 494. doi:10.1016/j.icarus.2014.07.010

\bibitem[Pasek et al.(2005)]{pa05} Pasek, M.~A., Milsom, J.~A., Ciesla, F.~J., et al.\ 2005, \icarus, 175, 1. doi:10.1016/j.icarus.2004.10.012

\bibitem[Rucska \& Wadsley(2023)]{Ru23} Rucska, J.~J. \& Wadsley, J.~W.\ 2023, \mnras, 526, 1757. doi:10.1093/mnras/stad2855

\bibitem[Schneeberger et al.(2023)]{Sc23} Schneeberger, A., Mousis, O., Aguichine, A., et al.\ 2023, \aap, 670, A28. doi:10.1051/0004-6361/202244670

\bibitem[Shakura \& Sunyaev(1973)]{sh73} Shakura, N.~I. \& Sunyaev, R.~A.\ 1973, X- and Gamma-Ray Astronomy, 55, 155

\bibitem[Smith et al.(1989)]{Sm89} Smith, B.~A., Soderblom, L.~A., Banfield, D., et al.\ 1989, Science, 246, 1422. doi:10.1126/science.246.4936.1422

\bibitem[Smith et al.(1986)]{Sm86} Smith, B.~A., Soderblom, L.~A., Beebe, R., et al.\ 1986, Science, 233, 43. doi:10.1126/science.233.4759.43

\bibitem[Stone \& Miner(1989)]{St89} Stone, E.~C. \& Miner, E.~D.\ 1989, Science, 246, 1417. doi:10.1126/science.246.4936.1417

\bibitem[Sromovsky et al.(2019)]{sr19} Sromovsky, L.~A., Karkoschka, E., Fry, P.~M., et al.\ 2019, \icarus, 317, 266. doi:10.1016/j.icarus.2018.06.026

\bibitem[Sromovsky et al.(2014)]{sr14} Sromovsky, L.~A., Karkoschka, E., Fry, P.~M., et al.\ 2014, \icarus, 238, 137. doi:10.1016/j.icarus.2014.05.016

\bibitem[Teanby et al.(2022)]{Te22} Teanby, N.~A., Irwin, P.~G.~J., Sylvestre, M., et al.\ 2022, \psj, 3, 96. doi:10.3847/PSJ/ac650f

\bibitem[Teanby et al.(2021)]{Te21} Teanby, N.~A., Gould, B., \& Irwin, P.~G.~J.\ 2021, \icarus, 354, 114045. doi:10.1016/j.icarus.2020.114045

\bibitem[Teanby et al.(2019)]{Te19} Teanby, N.~A., Irwin, P.~G.~J., \& Moses, J.~I.\ 2019, \icarus, 319, 86. doi:10.1016/j.icarus.2018.09.014

\bibitem[Teanby \& Irwin(2013)]{Te13} Teanby, N.~A. \& Irwin, P.~G.~J.\ 2013, \apjl, 775, L49. doi:10.1088/2041-8205/775/2/L49

\bibitem[Tsiganis et al.(2005)]{Ts05} Tsiganis, K., Gomes, R., Morbidelli, A., et al.\ 2005, \nat, 435, 459. doi:10.1038/nature03539

\bibitem[Tyler et al.(1986)]{Ty86} Tyler, G.~L., Sweetnam, D.~N., Anderson, J.~D., et al.\ 1986, Science, 233, 79. doi:10.1126/science.233.4759.79

\bibitem[Venot et al.(2020)]{Ve20} Venot, O., Cavali{\'e}, T., Bounaceur, R., et al.\ 2020, \aap, 634, A78. doi:10.1051/0004-6361/201936697


\bibitem[Weidenschilling(1997)]{we97} Weidenschilling, S.~J.\ 1997, \icarus, 127, 290. doi:10.1006/icar.1997.5712

\bibitem[Williams \& Cieza(2011)]{wi11} Williams, J.~P. \& Cieza, L.~A.\ 2011, \araa, 49, 67. doi:10.1146/annurev-astro-081710-102548


\bibitem[Zahnle et al.(2023)]{Za23} Zahnle, K.J. \ 2023, Uranus Flagship: Investigations and Instruments for Cross-Discipline Science Workshop 2023, LPI Contrib. No. 2808


\end{thebibliography}
\end{document}